\documentclass[prl,twocolumn,amsmath,amssymb,floatfix]{revtex4}
\usepackage{graphicx}% Include figure files

\newcommand{\beq}{\begin{equation}}
\newcommand{\eeq}{\end{equation}}
\newcommand{\beqa}{\begin{eqnarray}}
\newcommand{\eeqa}{\end{eqnarray}}

\newcommand{\q}{{\bf q}}
\newcommand{\kk}{{\bf k}}
\newcommand{\om}{\omega}

\newcommand{\be}{\begin{eqnarray}}
\newcommand{\ee}{\end{eqnarray}}
\newcommand{\bfr}{{\bf r}}

\newcommand{\bfk}{{\bf k}}

\newcommand{\bfQ}{{\bf Q}}

\newcommand{\wbe}{\begin{widetext}}
\newcommand{\wee}{\end{widetext}}

\begin{document}
\title{Fulde-­Ferrell-­Larkin­-Ovchinnikov state in bilayer dipolar systems}

\author{Hao Lee,$^{1,2,3}$ S. I. Matveenko,$^{4,3}$ Daw-Wei Wang,$^{1,2}$ and G. V. Shlyapnikov$^{3,5,6,7,8,9}$}

\affiliation{
$^{1}$ Physics Department, National Tsing Hua University, Hsinchu 30013, Taiwan\\
$^{2}$ Physics Division, National Center for Theoretical Sciences, Hsinchu 30013, Taiwan\\
$^{3}$ LPTMS, CNRS, Universit{\'e} Paris-Sud, Universit{\' e} Paris-Saclay, Orsay 91405, France\\
$^{4}$ L.D. Landau Institute for Theoretical Physics  RAS, Moscow,  Kosygin str. 2, 119334, Russia\\
$^{5}$ SPEC, CEA, CNRS, Universit{\' e} Paris-Saclay, CEA Saclay, Gif sur Yvette 91191, France\\
$^{6}$ Russian Quantum Center, Skolkovo, Moscow 143025, Russia\\
$^{7}$ Russian Quantum Center, National University of Science and Technology MISIS, Moscow 119049, Russia\\
$^{8}$ Van der Waals-Zeeman Institute, Institite of Physics, University of Amsterdam, Science Park 904, 1098 XH Amsterdam, The Netherlands\\
$^{9}$ Wuhan Institute of Physics and Mathematics, Chinese Academy of Sciences, 430071 Wuhan, China
}
%\PACS{67.80.bd,  67.85.Hj, 03.65.Vf}

\begin{abstract}
We study the
% ground state and finite temperature
phase diagram of fermionic polar molecules in a bilayer system,
with an imbalance of molecular densities of the layers. For the imbalance exceeding a critical value  the
system undergoes a transition from the uniform interlayer superfluid to the Fulde-Ferrell-Larkin-Ovchinnikov
(FFLO) state with mostly a stripe structure,
and at sufficiently large imbalance a transition from the FFLO to normal phase.
Compared to the case of contact interactions, the FFLO regime is enhanced by the long-range character of the interlayer dipolar interaction, which can combine the $s$-wave and $p$-wave pairing  in the order parameter.

\end{abstract}
\maketitle

%-------------------------------------------------------------------------------

%\section{Introduction: FFLO}
Exotic many-body quantum states in population imbalanced spin-1/2 Fermi systems attract a great deal of interest, to a large extent due to expected non-conventional transport properties.
Among these states, the most actively studied is the Fulde-Ferrell-Larkin-Ovchinnikov (FFLO) phase \cite{FF,LO}, in which Cooper pairs have finite momenta and the order parameter shows a lattice structure on top of a uniform background.
Theoretical studies of the FFLO phase in condensed matter are lasting for decades \cite{Bulaevski,Buzdin,BuzdinPLA97,Casalbuoni,FFLOHeavyF}, and rapid developments in the field of ultracold quantum gases stimulated the studies of this phase in two-component imbalanced Fermi gases \cite{LeoRPP, GubbelsPhysRep, CombescotEPL, KetterleRev,  LeoPRA84, MuellerPRA79, SheehyPRA85}.
However, experimental verification of the existence of the FFLO state is still in progress \cite{BianchiPRL03, RadovenNat03, Ketterle06, Ketterle07, HuletScience06, KetterleRev}.
In ultracold gases the search for the FFLO phase is actively pursued for strongly interacting fermions, where one has a crossover from Bardeen-Cooper-Shrieffer (BCS) superfluid to Bose-Einstein condensate of weakly bound molecules \cite{Pitaev, BlochRMP, Levin, Ketterle06, Ketterle07, HuletScience06}, and the population imbalance is expected to lead to the FFLO state \cite{MuellerPRA79, LeoPRA84, SheehyPRA85}.

%\section{Introduction: ultracold molecules}
Recent advances in creating ultracold polar molecules \cite{Ni, YeNJP} interacting with each other via long-range anisotropic dipole-dipole forces open fascinating prospects for many-body physics \cite{BaranovPhysRep, ZollerChemRev}.
A variety of novel many-body states was proposed for fermionic dipoles in two dimensions (2D) \cite{Taylor,Cooper2009,Gora,Pikovski2010,Potter,Sun,Miyakawa,DasSarmaFL,Ronen,BaranovSieb,
ZinnerDW,Parish,Babadi,Lu1,Lu2,Giorgini}, including interlayer superfluids with the BCS-BEC crossover in a bilayer %geometrey
geometry \cite{Pikovski2010,Ronen,ZinnerDW}.
Importantly, in 2D the decay of polar molecules due to ultracold chemical reactions \cite{Jin,Jin2} can be suppressed by orienting the dipoles perpendicularly to the plane, which induces a strong intermolecular repulsion \cite{Bohn1,BaranovM,YeM}.
Together with possible experiments with non-reactive polar molecules \cite{ZwierleinW,GrimmT}, this forms a promising path towards new many-body quantum states.  %In this paper

In this letter we  predict wide possibilities for creating the FFLO phase of polar molecules in a bilayer geometry,
with a finite imbalance of molecular densities of the layers. Cooper pairs are formed by molecules belonging to
different layers due to the interlayer   dipole-dipole interaction, and  the most favoured is the FFLO state with a
stripe structure of the order parameter.  Remarkably, the FFLO regime of this interlayer superfluid is
 enhanced by the long-range character of the dipolar interaction, which in an imbalanced system may lead to Cooper pairs representing superpositions of contributions of various partial waves.   Our work thus opens a new direction to investigate novel superfluids of
fermionic particles with population imbalance.

%========
\begin{figure}[htb]
\includegraphics[width=8.8cm]{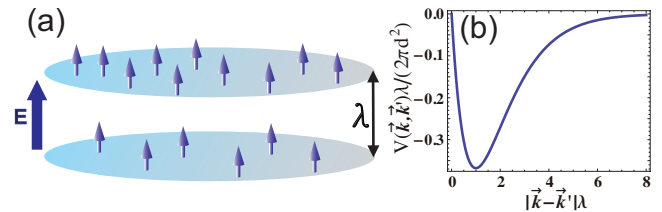}
\caption{(a)The bilayer system with population imbalance of polar molecules (see text).
 (b)The interlayer dipolar interaction in the k-space in units of $2\pi d^2/\lambda$ versus $|{\bf k}-{\bf k}'|$, as shown in Eq.(\ref{int0k}).}
\label{Scheme}
\end{figure}
%======
%%%%%%%%%%%%%%%%%%
{\it Interlayer interaction and order parameters:}
We consider identical fermionic polar molecules in a bilayer geometry, oriented perpendicularly to the
layers by an electric field (see Fig. 1).  The interlayer dipole-dipole interaction is partially attractive and it
may lead to interlayer superfluid pairing \cite{Pikovski2010,Potter,Ronen,ZinnerDW}. The inlayer interaction is
purely repulsive  and it only renormalizes the chemical potential  \cite{Pikovski2010, ZinnerDW}, so that below
 we omit this interaction. We assume that molecular densities in the layers are different from each other,
but in each layer the Fermi energy greatly exceeds  the binding energy of interlayer dimers (bound states
of dipoles belonging to different layers \cite{Pikovski2010, Ronen, Klawunn}).  Thus, the only effect of the interlayer dipole-dipole interaction will
 be the fermionic superfluid pairing, which we consider in the weakly interacting BCS regime.
% For simplicity, we assume the layer width is negligible comparing to the interlayer separation, $\lambda$.

 The interlayer dipole-dipole interaction potential and its Fourier transform are given by
\begin{eqnarray}
&V(\bfr)=d^{2}(r^{2}-2\lambda^{2})/(r^{2}+\lambda^{2})^{5/2}, \label{int0r} \\
&V_{\bfk\bfk'}= \int d\bfr V(\bfr)e^{i(\bfk'-\bfk)\cdot\bfr} =-2\pi d^{2}\kappa e^{-\kappa\lambda}, \label{int0k}
\end{eqnarray}
where $\kappa\equiv|\bfk-\bfk'|$, $d$ is the effective dipole moment of a molecule, and $\bfr$ is the inplane separation between  two dipoles.%,
% and $\lambda$ is the interlayer spacing.
The pairing potential can be expanded in a series in angular momenta $V_{\bfk\bfk'} = \sum _l V_l (k, k') e^{i(\phi - \phi')l}$. The leading part of the scattering (interaction) amplitude can be obtained in the Born approximation. It is important that in contrast to contact interactions, the dipolar (interlayer) interaction amplitude with $\l|>0$, and especially the $p$-wave amplitude, can be comparable with the $s$-wave amplitude. 

After omitting the inlayer interaction the system maps onto spin-1/2 fermions with intercomponent
dipolar  interaction. The Hamiltonian reads
  $H=H_0+H_I$, where $H_0=\sum_{\mathbf{k},\sigma}\xi_{\mathbf{k},\sigma} c^{\dagger}_{\mathbf{k}\sigma}c_{\mathbf{k}\sigma}$ is the kinetic energy, and
\begin{eqnarray}
H_I=\sum_{\mathbf{kk'q}}V_{\mathbf{kk'}}c^{\dagger}_{-\mathbf{k'}+\frac{\mathbf{q}}{2},\uparrow}
c^{\dagger}_{\mathbf{k'}+\frac{\mathbf{q}}{2},\downarrow}c_{\mathbf{k}+\frac{\mathbf{q}}{2},\downarrow}
c_{-\mathbf{k}+\frac{\mathbf{q}}{2},\uparrow}
\label{H_I}
\end{eqnarray}
%----
is the interaction energy. Here $c_{\mathbf{k}\sigma}$ are  fermionic field operators, $\sigma=\uparrow, \downarrow$, stands for the layer index,  $\xi_{\mathbf{k}\sigma}= k^{2}/2m-\mu_{\sigma}$ (hereinafter we put $\hbar=1$), and $\mu_{\sigma}=\mu\pm h$ are  chemical potentials of the layers. When the densities in the layers are not equal to each other,
there is a difference in the Fermi momenta of the two (pseudo)spin states,
 $\delta k_F\equiv |k_{F,\uparrow}-k_{F,\downarrow}|\neq 0$, and the effective magnetic field $h \neq 0$.

Relying on  Eq.(\ref{H_I}) we define the pairing gap
function as $\Delta_{\mathbf{kQ}}\equiv\sum_{\mathbf{k}'}V_{\mathbf{kk'}}\langle c_{\mathbf{k}'+\frac{\mathbf{Q}}{2},\downarrow}c_{-\mathbf{k}'+\frac{\mathbf{Q}}{2},\uparrow}\rangle$, where $\bfk'$ is the relative momentum of two paired fermions, and $\bfQ$ is their center-of-mass (CM) momentum. In the coordinate  space, the order parameter is then consisting of   pairing wavefunctions
 for several  CM momenta:
%---------
\be
\Delta_{\mathbf{k}}(\mathbf{R})=\sum ^{N_{Q}}_{n=1}\Delta_{\bfk\bfQ_n}e^{i\mathbf{Q}_n\cdot\mathbf{R}},
\label{Delta_k}
\ee
%-------
where $\mathbf{R}$ is the CM position of the Cooper pair, and $\mathbf{Q}_n$ is the CM momentum
 involved in the order parameter.
 Below we consider several symmetries of the order parameter:
  uniform superfluid ($N_Q=1$ and $\bfQ_1=0$), plane wave  FFLO  ($N_{Q}=1$ and  $\bfQ_{1}=q\hat{x}$,
   where $\hat{x}$ is a unit vector in the $x$ direction), stripe FFLO
   ($N_{Q}=2$ and $\bfQ_{1,2}=\pm q\hat{x}$),
   and  triangular state ($N_{Q}=3$ and three $\bfQ_n$ vectors have the same amplitude,
     with $2\pi/3$ difference in their orientation). Close to the FFLO-normal phase boundary we also consider square and hexagonal FFLO structures.

%%%%%%%%%%%%%%%
%-----
%\begin{figure}[htb]
%\includegraphics[width=9cm]{Gap.eps}
%\caption{The gaps $\Delta_{\mathbf{k}}(\mathbf{R}=0)$ of various superfluid states as functions of $\mathbf{k}$ in 2D at $T=0$, $k_{F}\lambda=0.4$, $r_{*}\equiv md^2/\hbar^2 = 0.5\; \lambda$, and $h/\mu=0.015$. }
%\label{Gap0}
%\end{figure}
%-------

%In order to obtain the  phase diagramm we use both numerical self-consistent solutions of Gor'kov equations  for the superfluid state , and  analitical results based on the Ginzburg-Landau functional near transition lines.

The  finite-temperature normal and anomalous Green  functions
 $G_{\sigma\sigma'}$ and
  $F^{\dagger}_{\sigma\sigma'}$  are found from the Gor'kov equations \cite{Gorkov,MahanMineev} (see Supplemental Material for details):
\begin{eqnarray}
&&G_{\sigma\sigma'}(\bfk_1,\bfk_2;i\omega_{n})
=\delta_{\sigma\sigma'}\delta_{\bfk_1,\bfk_2}\nonumber\\
&\times &\left(i\omega_{n}-\xi_{\mathbf{k}_1\sigma}-\sum^{N_Q}_{m=1}\frac{\Delta_{\bfk_1,\bfQ_m}\Delta^{\dagger}_{\bfk_2-\bfQ_m,\bfQ_m}}{i\omega_{n}+\xi_{\bfk_1-\bfQ_m,\sigma}}\right)^{-1};
\label{Gkk}
\end{eqnarray}
\begin{eqnarray}
&&F^{\dagger}_{\sigma\sigma'}(\bfk_1,\bfk_2;i\omega_{n})
=\frac{-\sum^{N_Q}_{m=1}\Delta^{\dagger}_{\bfk_1,\bfQ_m}\delta_{\bfk_1+\bfQ_m,\bfk_2}}{i\omega_{n}+\xi_{\mathbf{-k}_1\sigma}}\\
&\times &G_{\sigma'\sigma'}(\bfk_2,\bfk_2;i\omega_{n})(1-\delta_{\sigma\sigma'}),\nonumber
\label{Fd}
\end{eqnarray}
where $\omega_n = (2n + 1)\pi T$ are Matsubara frequencies, and $T$ is the temperature.
The gap equation can then be obtained self-consistently:
\begin{equation}
\Delta^{*}_{\bfk\bfQ}=-T\sum_{n,\mathbf{k'}}V_{\mathbf{kk'}}F^{\dagger}_{\uparrow\downarrow}
(\mathbf{k'}-\frac{\bfQ}{2},\mathbf{k'}+\frac{\bfQ}{2};\omega_{n}).
 \label{gap}
 \end{equation}
We thus identify the Gor'kov equations for the Green functions and the gap equations for $\Delta^*_{{\bf k}{\bf Q}}$ and $\Delta_{{\bf k}{\bf Q}}$ as self-consistent Gor'kov equations.

The plane wave, stripe, triangular, square, and hexagonal phases break the rotational symmetry and have an anisotropic gap in the momentum space. In principle, such gap function can be
 measured using Bragg spectroscopy by exciting particles with a finite momentum.

 In the following, we  first solve the self-consistent Gor'kov equations by assuming a certain gap function symmetry (uniform, plane wave, stripe, or triangular) and then determine the  phase diagram by comparing the obtained
 free energies  of these candidates. The derivation of equations (\ref{Gkk})-(\ref{F4}) and a detailed presentation of the numerical procedure of obtaining the phase diagrams at zero and finite temperatures are contained in the Supplemental Material.

Near the phase transition line  (superfluid -- normal state), where  the order parameter is small,
we  may use the Ginzburg-Landau  free energy functional $F = F_2 + F_4+...$, with
\begin{widetext}
\beq
F_2 = \sum \Delta_{\kk\q}(V^{-1})_{\kk \kk'} \Delta^*_{\kk'\q} - T\sum  |\Delta_{\kk\q}|^2 G_+ (\kk+\frac{\q}{2}, \om_n)
 G_- (-\kk +\frac{\q}{2}, -\om_n), \label{F2}
 \eeq
 \beq
 F_4 = \frac{T}{2}\sum_{\{\q_j\},  \kk n}\Delta^*_{\kk\q_1}\Delta^*_{\kk\q_2}\Delta_{\kk\q_3}\Delta_{\kk\q_4}
  G_+ (\kk+\q_1, \om_n)
 G_- (-\kk +\q_3 -\q_1, -\om_n)
 G_+ (\kk+\q_4, \om_n)
 G_- (-\kk, -\om_n) \delta_{ \q_1 +\q_2 , \q_3 +\q_4}, \label{F4}
 \eeq
 \end{widetext}
where the Green's functions of the normal state are
\beq
G_{\pm} (\kk, \om_n) = \frac{1}{i\om_n - \xi_{\kk} \mp h},
\eeq
 and  $V^{-1}$ is the inverse matrix of $V _{{\bf k}{\bf k}'}$.
The  term $F_4$  is necessary to find the minimum energy configuration, whereas $F_2$
determines the tricritical point.

%%%%%%%%%%%%
%=======
\begin{figure}[htb]
\includegraphics[width=8cm]{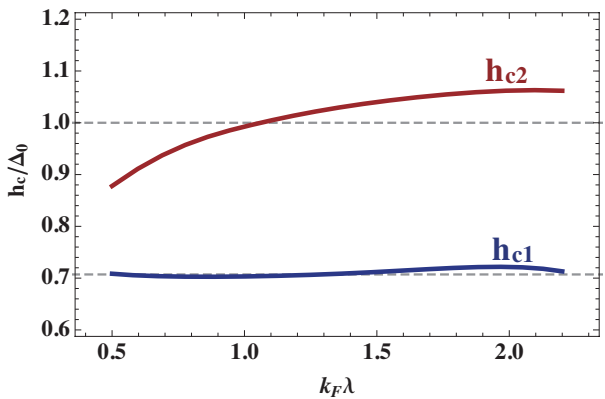}
\caption{Critical fields $h_{c1}$ and $h_{c2}$ at $T=0$ in units of $\Delta_0$ versus the parameter $k_F \lambda$ at $k_Fr_*=0.7$, where $r_*=md^2$ is the dipole-dipole distance.  The dashed lines show $h_{c1}$ and $h_{c2}$ for the case of contact interaction.
}
\label{0Tphase}
\end{figure}
%========
{\it Zero temperature:}
At zero field ($h=0$) the ground state is a uniform superfluid with the order parameter on the Fermi surface $\Delta_0(k_F)\equiv\Delta_0$, where $k_F=\sqrt{2m\mu}$ is the Fermi momentum at $h=0$.
For a given interaction strength the FFLO phase emerges at a critical value $h_{c1}$, and our calculations show that this is a stripe phase. For sufficiently large field $h_{c2}$ the ground state becomes normal. Note that the stripe FFLO phase is clearly the ground state at $h$ that are lower than $h_{c2}$ by a few percent (see Supplemental Material). The dependence of $h_{c1}$, $h_{c2}$ on the parameter $k_F \lambda$ displayed in Fig.~2 shows that for $k_F\lambda>1$ the FFLO region is significantly wider than in the case of contact interaction, where the FFLO phase emerges for $\Delta_0>h>\Delta_0/\sqrt{2}\approx 0.707\Delta_0$. Here we arrive at a very important point. For $k_F\lambda<1.055$ the $p$-wave interaction on the Fermi surface is repulsive and interlayer Cooper pairs only contain the contribution of an attractive $s$-wave interaction. However, if $k_F\lambda>1.055$, then the $p$-wave interaction $ V_1(k_F,k_F)$ becomes also attractive and Copper pairs are already composed of both $s$-wave and $p$-wave contributions. This makes the modulus of the order parameter larger and requires a higher field $h_{c2}$ to destroy superfluidity and get to the normal state (see Supplemental Material).

%%%%%%%%%%%%%%%%%%
\begin{figure*}[htbp]
\includegraphics[width=15cm]{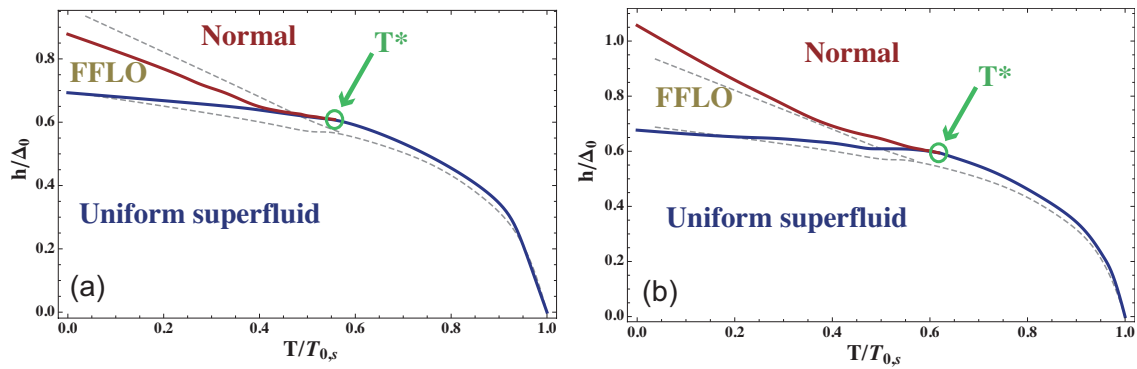}
\caption{(a) Finite temperature phase diagram in terms of   $T/T_{0,s}$  and the imbalance
 $h/\Delta_0$. In (a) $k_F\lambda =0.5$ and $r_*=\lambda/2$, and in (b) $k_F\lambda =2.2$ and $r_*=\lambda$. The dashed curves are the phase boundaries for the case of contact interaction.
}
\label{fTphase}
\end{figure*}
%%%%%%%%%%%%%%%%%%
{\it Finite temperature phase diagram:}
At finite temperatures the equilibrium phases are determined by comparing the free energies of the
uniform superfluid, FFLO, and the normal state.
In Fig. \ref{fTphase} the phase diagram is presented in terms of $T/T_{0,s}$ and $h/\Delta_0$, where $T_{0,s}$
is the transition temperature at $h=0$ (provided by the $s$-wave pairing).
With increasing temperature,  the critical field $h_{c1}$ for the transition from the uniform
superfluid  to FFLO phase  decreases. So does the imbalance  for the transition from the FFLO to
the normal state. The three phases (SF, FFLO, and normal) merge  at the tricritical point $T^*$. For $k_F\lambda<1$ we have $T^*\approx 0.56 T_{0,s}$
like in the case of contact interactions. However, for $k_F\lambda>1$ the $p$-wave pairing contribution to the order parameter comes into play and the ratio $T^*/T_{0,s}$ increases.
Already for $k_f\lambda=2.2$ we obtain $T^*\simeq 0.62 T_{0,s}$ and somewhat wider FFLO region than in the case of contact interactions (see Fig. \ref{fTphase}).

The  FFLO phase is mostly the Larkin-Ovchinnikov stripe state. 
In the temperature interval from $0.02T_{0,s}$ to $0.3T_{0,s}$ the stripe phase has the lowest free energy 
%for $h$ lower by more than $5\%$ than $h_{c2}$.
for $h$ lower than $h_{c2}$ by more than $5\%$.
For $h$ closer to $h_{c2}$ our numerical calculations based on the Ginzburg-Landau functional show that the equilibrium state is a triangular FFLO (it is also recovered from the self-consistent Gor'kov equations at $h$ close to $0.95h_{c2}$), which with decreasing temperature becomes a square and then hexagonal FFLO (see Supplemental Material). This sequence of FFLO states is similar to that observed for contact interactions \cite{Shimahara,FFLOHeavyF,CombescotEPL}. 

The structure of the phase diagram is similar to that
in the case of contact interactions \cite{LO,BuzdinPLA97,Shimahara}. However, in our case the FFLO region
significantly  depends on the parameters,  so that  $T^*/T_{0,s}$  and $h^*/\Delta_0$ are not universal numbers as they are for  contact interactions
where $T^* \approx 0.56 T_{0,s}$ and the corresponding critical imbalance is $h^* \approx 0.6 \Delta_0$ \cite{BuzdinPLA97,Shimahara}).

The tricritical temperature can be determined analytically (see Supplemental Material for details):
\begin{equation}
 -\ln\frac{T_{0,s}}{T_{0p}}+{\rm Re}\left[\Psi\left(\frac{1}{2}+i\frac{h^*}{2\pi T^*}\right)-\Psi\left(\frac{1}{2}\right)\right]=0;   \label{Tcanal1}
 \end{equation}
 \begin{equation}
 -{\rm Re}\left[\Psi''  \left( \frac{1}{2} + i\frac{h^*  }{2\pi T^*}\right)\right] = \frac{\left({\rm Im} \Psi' \left(1/2 + ih^*/2\pi T^*\right)\right)^2}{\ln(T_{0,s}/T_{0,p})},   \label{Tcanal}
 \end{equation}
where $T_{0,p}$ is the critical temperature of superfluid transition for the $p$-wave pairing at $h=0$, and $\Psi$  is the digamma function. In the limit $T_{0,p}\rightarrow 0$ Eqs (\ref{Tcanal1}) and (\ref{Tcanal}) give the known result for the contact interaction, which is specified in the previous paragraph. These equations also reproduce numerical results of Figs. 3 and 4.
The maximum tricritical temperature $T^* \to  T_{0,s}$ is reached in the limit $k_F \lambda \gg 1$, where
$T_{0,p}$ is close to $T_{0,s}$. 
However, the critical temperature $T_{0,s}$ decreases with increasing $k_F\lambda$. For $k_F\lambda\simeq 2$, where one can still hope to achieve $T_{0,s}$ on the level of nanokelvins  (see below), we obtain $T^* \approx 0.62 T_{0,s}$.

The lattice wavevector $|{\bf Q}_m|$  behaves as  $Q\sim \sqrt{T_c^* -T}$ near the tricritical point and vanishes at $T=T_c^*$ (see Supplemental Material).

As is already said above, the equilibrium FFLO phase in our case is the stripe state. Like in the case of contact interactions \cite{burkh,Shimahara}, the transition from the uniform superfluid to FFLO state is of the first order, whereas the transition from the FFLO to normal phase is of the second order. At 2D densities $\sim10^9\text{cm}^{-2}$ the Fermi energy for weakly reactive NaLi molecules or non-reactive NaK molecules is $\epsilon_F\sim 1 \mu$K, and for the interlayer spacing $200$ nm we have $k_F\lambda\simeq 2$. Then, on approach to the strongly interacting regime with $k_Fr_*$ slightly exceeding unity the superfluid transition temperature will be up to $10$ nK. Larger values of $k_F\lambda$ and, hence, higher ratios $T^*/T_{0,s}$ would require $k_Fr_*$ significantly larger than unity, so that the inlayer interaction becomes very strong driving the system far away from the dilute regime.  This case is beyond the scope of the present paper.

%\begin{figure}[htb]
%\includegraphics[width=8cm]{PhaseDiagFTkfl05.eps}
%\caption{Finite temperature phase diagram in terms of   $T/T_{0,s}$  and the imbalance
% $h/\Delta_0$  at
% $k_F\lambda =0.5$ and $r_*=\lambda/2$.
%}
%\label{fTphase}
%\end{figure}
%==========
%\begin{figure}[htb]
%\includegraphics[width=8cm]{PhaseDiagFTkfl22.eps}
%\caption{Finite temperature phase diagram in terms of   $T/T_{0,s}$  and the imbalance
% $h/\Delta_0$ at
% % $k_Fr_*=0.25$ and
% $k_F\lambda =2.2$ and $r_*=\lambda$.
% }
%\label{fTphase2}
%\end{figure}

%\begin{figure}[htb]
%\includegraphics[width=6cm]{PhaseDiagFTkfl22-5.eps}
%\caption{{\bf Wave vector  {\bf q} near tricritical point.}  The optimal $\bar{q} \equiv q  k_F /(2mh) \propto \sqrt{1 - T/T^*}$
%as $T \to T^*$. }
%\label{qqq}
%\end{figure}

{\it Conclusions:}  We used  both a theoretical field approach based on the Gor'kov equations and the theory of phase transitions  based on the Ginzburg-Landau free energy functional to study  possible FFLO phases
 in a bilayer system of fermionic polar molecules
% with finite density imbalance.
with a finite imbalance of molecular densities of the layers.
 %The equilibrium FFLO phase is always triangular at both zero and finite temperatures and it exists in a much
 % wider parameter region than in the case of contact interactions.
   Our work demonstrates the importance of the long-range character of the dipole-dipole interaction, which can combine the $s$-wave and $p$-wave pairing in the order parameter
and enhance the FFLO regime.
   The  observation of this FFLO state is feasible
  for non-reactive NaK molecules or weakly reactive NaLi molecules at temperatures $\sim10\text{nK}$.
%   and 2D densities approaching $10^{9} cm^{-2}$.

We are grateful to Alexander Buzdin for fruitful discussions. We acknowledge support from MOST and NCTS in Taiwan. The research leading to these results
has received funding from the European Research Council under European Community's Seventh Framework Programme (FP7/2007-2013 Grant Agreement no. 341197).
%%%%%%%%%%%%%%%%%%%%%%%%%%%%%%%%%%%%

%%%%%%%%%%%%%%%%%%%%%%%%%%%%%%%%%%%%
%\bibliographystyle{apsrev.bst} 　
%\bibliography{ref1}

\end{document}

%\title{Supplementary Material for\\ ``Population-imbalanced fermionic polar molecules in bilayer system''}
\title{Supplemental Material for\\ Fulde-­Ferrell-­Larkin­-Ovchinnikov state in bilayer dipolar systems}

\author{Hao Lee,$^{1,2,3}$ S. I. Matveenko,$^{4,3}$ Daw-Wei Wang,$^{1,2}$ and G. V. Shlyapnikov$^{3,5,6,7,8,9}$}

\affiliation{
$^{1}$ Physics Department, National Tsing Hua University, Hsinchu 30013, Taiwan\\
$^{2}$ Physics Division, National Center for Theoretical Sciences, Hsinchu 30013, Taiwan\\
$^{3}$ LPTMS, CNRS, Universit{\'e} Paris-Sud, Universit{\' e} Paris-Saclay, Orsay 91405, France\\
$^{4}$ L.D. Landau Institute for Theoretical Physics  RAS, Moscow,  Kosygin str. 2, 119334, Russia\\
$^{5}$ SPEC, CEA, CNRS, Universit{\' e} Paris-Saclay, CEA Saclay, Gif sur Yvette 91191, France\\
$^{6}$ Russian Quantum Center, Skolkovo, Moscow 143025, Russia\\
$^{7}$ Russian Quantum Center, National University of Science and Technology MISIS, Moscow 119049, Russia\\
$^{8}$ Van der Waals-Zeeman Institute, Institite of Physics, University of Amsterdam, Science Park 904, 1098 XH Amsterdam, The Netherlands\\
$^{9}$ Wuhan Institute of Physics and Mathematics, Chinese Academy of Sciences, 430071 Wuhan, China
}
\date{\today}
\maketitle

In the %Supplenental
Supplemental Material we present the derivation of the solutions of the Gor'kov equations for the Green functions
$G_{\alpha,\beta}(\bfk, \bfk', \tau)$, $ F_{\alpha,\beta}(\bfk, \bfk', \tau)$, $ F^{\dagger}_{\alpha,\beta}(\bfk, \bfk', \tau)$,
given by Eqs.~(4) and (5) in the main text.
We then present the self-consistent gap equation in terms of these Green functions as given by Eq.~(6)
of the main text.
In the next part of the Supplemental Material we discuss the Ginzburg-Landau approach, which includes the $s$- and $p$-wave pairing. We also present the derivation of the critical magnetic field and  the tricritical temperature.
The last part of the Supplemental material is dedicated to a detailed description of our numerical procedure for the calculation of the free energy for the uniform superfluid, FFLO states, and normal state.
\vspace{0.3cm}

\section{{{\bf Solutions  of the  Gor'kov  equations  for the Green functions $G$, $F$, $F^{\dagger}$
and a self-consistent gap equation}}}
%\vspace{0.2cm}

We start with the Hamiltonian $H=H_0+H_I$, where

\begin{equation} \label{H0}
H_0=\sum_{\mathbf{k}\sigma}\xi_{\mathbf{k}\sigma}c^{\dagger}_{\mathbf{k}\sigma}c_{\mathbf{k}\sigma},
\end{equation}
with  $\xi_{\bfk\sigma}\equiv\epsilon_k-\mu_\sigma$, and $H_{I}$ describes the interaction between particles:
\begin{eqnarray}
H_I=\sum_{\mathbf{kk'}m}V_{\mathbf{kk'}}c^{\dagger}_{-\mathbf{k'}+\frac{\mathbf{Q}_m}{2},\uparrow}c^{\dagger}_{\mathbf{k'}+\frac{\mathbf{Q}_m}{2},\downarrow}c_{\mathbf{k}+\frac{\mathbf{Q}_m}{2},\downarrow}c_{-\mathbf{k}+\frac{\mathbf{Q}_m}{2},\uparrow},
\label{Hori}
\end{eqnarray}
and we keep the same notations as in the main text.
The Green functions are defined  as \cite{Gorkov}
\begin{eqnarray}
G_{\alpha\beta}(\mathbf{k},\tau_{1};\mathbf{k'},\tau_{2})&=&-\left\langle T_{\tau}c_{\mathbf{k}\alpha}(\tau_{1})c^{\dagger}_{\mathbf{k'}\beta}(\tau_{2}) \right\rangle,\\
F_{\alpha\beta}(\mathbf{k},\tau_{1};\mathbf{k'},\tau_{2})&=&\left\langle T_{\tau}c_{\mathbf{k}\alpha}(\tau_{1})c_{-\mathbf{k'}\beta}(\tau_{2}) \right\rangle,\\
F^{\dagger}_{\alpha\beta}(\mathbf{k},\tau_{1};\mathbf{k'},\tau_{2})&=&\left\langle T_{\tau}c^{\dagger}_{-\mathbf{k}\alpha}(\tau_{1})c^{\dagger}_{\mathbf{k'}\beta}(\tau_{2}) \right\rangle,
\label{defGF}
\end{eqnarray}
where
% $\tau\equiv 1/T$ (here we define $k_{B}=1$),
$T_{\tau}$ is the imaginary  time-ordering operator, and $c_{\mathbf{k}\alpha}(\tau)\equiv e^{\tau H}c_{\mathbf{k}\alpha}e^{-\tau H}$.
In the momentum-frequency representation we obtain
\begin{eqnarray}
G_{\alpha\beta}(\mathbf{k},\mathbf{k'};\tau)=T\sum_{n}G_{\alpha\beta}(\mathbf{k},\mathbf{k'},\omega_{n})
e^{-i\omega_{n}\tau},
\end{eqnarray}
and similar expressions for $F$, $F^{\dagger}$,
where $\tau=\tau_{1}-\tau_{2}$ and $\omega_{n}=(2n+1)\pi T$ is the Matsubara frequency for fermions.
%%%%%%%%%%%%%add the derivation here%%%%%%%%%%%%%%%%%%%
Using  equations of motion for the operators $c_{\mathbf{k}\sigma}$:
\begin{eqnarray}
\frac{\partial}{\partial\tau}c_{\mathbf{k}\sigma}(\tau)&=&\left[H(\tau),c_{\mathbf{k}\sigma}(\tau)\right]\nonumber\\
&=&-\xi_{\bfk\sigma}c_{\bfk\sigma}(\tau)-\sum_{\bfk_1\bfk'm}V_{\bfk_1\bfk'}c^\dagger_{\bfk'+\bfQ_m/2,\sigma'}
(\tau)c_{\bfk_1+\bfQ_m/2,\sigma'}(\tau)c_{-\bfk_1+\bfQ_m/2,\sigma}(\tau),
\end{eqnarray}
and the definition of $c_{\mathbf{k}\sigma}(\tau)$, we obtain  for the derivative of the Green function:
\begin{eqnarray}
\frac{\partial G_{\sigma\sigma'}(\mathbf{k,k'};\tau)}{\partial\tau}&=&-\left\langle T_{\tau}\left[H(\tau),c_{\mathbf{k},\sigma}(\tau)\right]c^{\dagger}_{\mathbf{k'},\sigma'}(0)\right\rangle-\delta(\tau)\delta(\mathbf{k-k'})\delta_{\sigma\sigma'}
\end{eqnarray}
\begin{eqnarray}
=&-&\sum_{\mathbf{k}_{1}}\left\langle \mathbf{k}\alpha |\xi_{\bfk_1\gamma}|\mathbf{k}_{1}\gamma\right\rangle G_{\gamma\beta}(\mathbf{k}_{1},\mathbf{k'})-\delta(\tau)\delta(\mathbf{k-k'})\delta_{\alpha\beta}\nonumber\\
&-&\sum_{\mathbf{k}_{1}m}V_{\mathbf{k'k}_{1}}\left\langle T_{\tau}c^{\dagger}_{\mathbf{-k+Q}_m,\gamma_{1}}(\tau)c_{\mathbf{-k}_{1}+\mathbf{Q}_m/2,\gamma_{2}}(\tau)c_{\mathbf{-k}_{1}+\mathbf{Q}_m/2,\gamma_{3}}(\tau)c^{\dagger}_{\mathbf{k'},\beta}(0)\right\rangle,
\label{eom}
\end{eqnarray}
where % $H_{0}=\sum_{\mathbf{k}\sigma}\xi_{\mathbf{k}\sigma}c^{\dagger}_{\mathbf{k}\sigma}c_{\mathbf{k}\sigma}$ is the kinetic energy term in the Hamiltonian and
$\alpha$, $\beta$, $\gamma$, $\gamma_{1}$, $\gamma_{2}$ and $\gamma_{3}$ are layer indices $\uparrow$ or $\downarrow$.
Following Ref. \cite{Gorkov} we use the mean-field approximation
 for the average of the last term of Eq.(\ref{eom}):
\begin{eqnarray}
&&\left\langle T_{\tau}c^{\dagger}_{\mathbf{-k+Q}_m,\gamma_{1}}(\tau)c_{\mathbf{-k}_{1}+\mathbf{Q}_m/2,\gamma_{2}}(\tau)c_{\mathbf{-k}_{1}+\mathbf{Q}_m/2,\gamma_{3}}(\tau)c^{\dagger}_{\mathbf{k'},\beta}(0)\right\rangle\nonumber\\
&\rightarrow &\left\langle T_{\tau}c^{\dagger}_{\mathbf{-k+Q}_m,\gamma_{1}}(\tau)c^{\dagger}_{\mathbf{k'}\beta}(0)\right\rangle\left\langle T_{\tau}c_{\mathbf{k}_{1}+\mathbf{Q}_m/2,\gamma_{2}}(\tau)c_{\mathbf{-k}_{1}+\mathbf{Q}_m/2,\gamma_{3}}(\tau)\right\rangle.
\end{eqnarray}

We then introduce the order parameter  as
\begin{eqnarray}
\Delta_{\alpha\beta}(\mathbf{k},\mathbf{Q}_m)=-T\sum_{\mathbf{k'},n}V_{\mathbf{k,k'}}F_{\alpha\beta}\left(\mathbf{k'}+\frac{\mathbf{Q}_m}{2},\mathbf{k'}-\frac{\mathbf{Q}_m}{2};\omega_{n}\right);
\label{gapF}
\end{eqnarray}

\begin{equation}
\Delta^{\dagger}_{\alpha\beta}(\mathbf{k},\mathbf{Q}_m)=-T\sum_{\mathbf{k'},n}V_{\mathbf{k,k'}}F^{\dagger}_{\alpha\beta}\left(\mathbf{k'}-\frac{\mathbf{Q}_m}{2},\mathbf{k'}+\frac{\mathbf{Q}_m}{2};\omega_{n}\right).
\end{equation}
Then the equation for the Green function $G$ acquires the form
%\begin{widetext}
\begin{equation}
\sum_{\mathbf{k}_{1}}\left\langle \mathbf{k}\alpha |(i\omega_{n}-\xi_{\mathbf{k}\alpha})|\mathbf{k}_{1}\gamma\right\rangle G_{\gamma\beta}(\mathbf{k}_{1},\mathbf{k'};\omega_{n})=-\sum_{m}\Delta_{\alpha\gamma}(\mathbf{k},\mathbf{Q}_m)
F^{\dagger}(\mathbf{k-Q}_m,\mathbf{k'};\omega_{n})+\delta(\mathbf{k-k'})\delta_{\alpha\beta}.
\label{g1}
\end{equation}
Similarly, we obtain  equations for the  anomalous Green functions:
\begin{equation}
\sum_{\mathbf{k}_{1}}\left\langle -\mathbf{k}_{1}\gamma |(i\omega_{n}+\xi_{\mathbf{-k}\alpha})| -\mathbf{k}\alpha\right\rangle F^{\dagger}_{\gamma\beta}(\mathbf{k}_{1},\mathbf{k'};\omega_{n})=-\sum_{m}\Delta^{\dagger}_{\alpha\gamma}(\mathbf{k,Q}_m)G_{\gamma\beta}(\mathbf{k+Q}_m,\bfk';\omega_{n}),
\label{g2}
\end{equation}
\begin{equation}
\sum_{\mathbf{k}_{1}}\left\langle \mathbf{k}\alpha |(i\omega_{n}-\xi_{\mathbf{k}_1\gamma})| \mathbf{k}_1\gamma\right\rangle F_{\gamma\beta}(\mathbf{k}_{1},\mathbf{k'};\omega_{n})=\sum_{m}\Delta_{\alpha\gamma}(\mathbf{k,Q}_m)G_{\beta \gamma}(\mathbf{-k'}, -\bfk+ {\bf Q}_m;-\omega_{n}).
\label{g3}
\end{equation}

The derived  Gor'kov equations (13) - (15)  are then solved using  the  self-consistent equations (11), (12). We thus identify Eqs.~(11)-(15) as self-consistent Gor'kov equations.
 The non-zero components of the order parameter are $\Delta_{\uparrow\downarrow}=-\Delta_{\downarrow\uparrow}$, and  below we use a simplified notation  $\Delta_{\uparrow\downarrow}(\mathbf{k},\mathbf{Q}_m)=\Delta_{\bfk,\bfQ_m}$.  We then write the solutions of  equations (\ref{g1}) - (\ref{g3}) in the form
\begin{eqnarray}
%G_{\sigma\sigma}(\mathbf{k,k};\omega_{n})&=&\frac{1}{(i\omega_{n}-\xi_{\mathbf{k}\sigma})-\sum_{\mathbf{q'}}\frac{\Delta^{2}}{i\omega_{n}+\xi_{\mathbf{k-q'}\sigma'}}},\\
%F^{\dagger}_{\alpha\beta}(\mathbf{k,k'};\omega_{n})&=&\frac{-1}{i\omega_{n}+\xi_{\mathbf{-k}\alpha}}\times\frac{\Delta^{\dagger}(\mathbf{k,q})}{(i\omega_{n}-\xi_{\mathbf{k'}\beta})-\sum_{\mathbf{q'}}\frac{\Delta(\mathbf{k',q'})\Delta^{\dagger}(\mathbf{k'-q',q'})}{i\omega_{n}+\xi_{\mathbf{k'-q'},\alpha}}}.
G_{\sigma\sigma'}(\bfk,\bfk';i\omega_{n})
&=&\frac{\delta_{\sigma\sigma'}\delta_{\bfk,\bfk'}}{i\omega_{n}-\xi_{\mathbf{k}\sigma}-\sum_{m}\frac{\Delta_{\bfk,\bfQ_m}\Delta^{\dagger}_{\bfk'-\bfQ_m,\bfQ_m}}{i\omega_{n}+\xi_{\bfk-\bfQ_m,\sigma}}},
\label{Gkk}
\\
F^{\dagger}_{\sigma\sigma'}(\bfk,\bfk';i\omega_{n})
&=&\frac{-\sum_{m}\Delta^{\dagger}_{\bfk,\bfQ_m}\delta_{\bfk+\bfQ_m,\bfk'}}{i\omega_{n}+\xi_{\mathbf{-k}\sigma}}\times G_{\sigma'\sigma'}(\bfk',\bfk';i\omega_{n})(1-\delta_{\sigma\sigma'})
\label{Fd}
\\
F_{\sigma\sigma'}(\bfk,\bfk';i\omega_{n})
&=&\frac{\sum_{m}\Delta_{\bfk,\bfQ_m}\delta_{\bfk-\bfQ_m,\bfk'}}{i\omega_{n}-\xi_{\mathbf{k}\sigma}}\times G_{\sigma'\sigma'}(-\bfk',-\bfk';-i\omega_{n})(1-\delta_{\sigma\sigma'})
\end{eqnarray}

The quasiparticle  spectrum $E_{{\bf k},{\bf Q},\sigma}$ is determined by the poles of the retarded Green function which is obtained by
analytical continuation of the temperature Green function (\ref{Gkk})  to the upper half-plane. This yields:
\begin{eqnarray} \label{qpspectrum}
E_{\bfk,{\bf Q},\sigma}=\xi_{\mathbf{k}\sigma}+\sum^{N_Q}_{m=1}\frac{\Delta_{\bfk,\bfQ_m}\Delta^{\dagger}_{\bfk-\bfQ_m,\bfQ_m}}{E_{\bfk,\sigma}+\xi_{\bfk-\bfQ_m,\sigma}}.
\label{exi}
\end{eqnarray}
%====
Equation (\ref{qpspectrum})  reduces to the conventional result $E_{\bfk} = \sqrt{ \xi_{\mathbf{k}}^2 + |\Delta_{\bfk}|^2}$
for a uniform superfluid.

We then calculate the  free energy  $F = E - TS$, where $E$
is the energy and $S$ is the entropy of  a system given by the usual relation:
\begin{equation}    \label{S}
S = -\sum_{\bfk, \sigma} [ f(E_{\bfk,\bfQ,\sigma}) \ln f(E_{\bfk,\bfQ,\sigma})  + (1-f(E_{\bfk,\bfQ,\sigma}) )\ln(1-f(E_{\bfk,\bfQ,\sigma}))],
\end{equation}
with $f(E_{\bfk,\bfQ,\sigma})$ being the distribution function. The energy is obtained in the mean-field BCS-type approach, which assumes that the order parameter is much smaller than the Fermi energy:
\beq    \label{E}
E=\left\langle H\right\rangle
=\sum_{\sigma\mathbf{k}}\xi_{\mathbf{k}\sigma}\left\langle c^{\dagger}_{\mathbf{k}\sigma}c_{\mathbf{k}\sigma}\right\rangle +\sum_{\mathbf{k},m}\Delta_{\mathbf{k,Q}_m}\left\langle c^{\dagger}_{-\mathbf{k}+\frac{\mathbf{Q}_m}{2},\uparrow}c^{\dagger}_{\mathbf{k}+\frac{\mathbf{Q}_m}{2},\downarrow}\right\rangle,
\eeq
where $\left\langle c^{\dagger}_{\mathbf{k}\sigma}c_{\mathbf{k}\sigma}\right\rangle=\frac{1}{2}+T\sum_{n}G_{\sigma\sigma}(\mathbf{k,k};\omega_{n})$ and $\left\langle c^{\dagger}_{-\mathbf{k}+\frac{\mathbf{Q}_m}{2},\uparrow}c^{\dagger}_{\mathbf{k}+\frac{\mathbf{Q}_m}{2},\downarrow}\right\rangle
=-T\sum_{n}F^{\dagger}_{\uparrow\downarrow}(\mathbf{k}-\frac{\mathbf{Q}_m}{2},\mathbf{k}+\frac{\mathbf{Q}_m}{2};\omega_{n})$.

In certain special cases we have explicit expressions. For  a uniform BCS superfluid phase,  ($|\bfQ_m|=0$), the energy acquires the form:
\begin{equation}  \label{EBCS}
E_{BCS}=\sum_{\bfk}\left(\xi_{k}-\frac{\xi^{2}_k}{E_{\bfk}}\tanh\frac{E_{\bfk}}{2T}-\frac{|\Delta_\bfk|^{2}}{2E_\bfk}\tanh\frac{E_\bfk}{2T}\right),
\end{equation}
where $E_\bfk=\sqrt{\xi^{2}_k+|\Delta_\bfk|^{2}}$. For the energy of the Fulde-Ferrell plane-wave phase ($N_Q=1$ and ${\bf Q}_1=Q\hat x$, where $\hat x$ is a unit vector in the $x$-direction) we obtain:
\begin{equation}   \label{EFF}
E_{FF}=\sum_{\bfk}\left(\mathcal{E}_{k,Q}-\frac{\mathcal{E}_{k,Q}^{2}}
{E_{\bfk,\bfQ}}\left(\frac{1}{2}\sum_{\sigma}\tanh\frac{E_{\bfk,\bfQ,\sigma}}{2T}\right)-\frac{|\Delta_{\bfk,\bfQ}|^{2}}{2E_{\bfk,\bfQ}}
\left(\frac{1}{2}\sum_{\sigma}\tanh\frac{E_{\bfk,\bfQ,\sigma}}{2T}\right)
+\sum_{\sigma}\frac{\sigma}{2}\left(h+\frac{\bfk\cdot\bfQ}{2m}\right)\tanh\frac{E_{\bfk,\bfQ,\sigma}}{2T}\right),
\end{equation}
%%%%%%%%%%%%
where $\mathcal{E}_{k,Q}=\xi_k^{2}+\frac{Q^{2}}{8m}$, $E_{\bfk,\bfQ}=\sqrt{\mathcal{E}_{k,Q}^{2}+|\Delta_{\bfk,\bfQ}|^{2}}$, and $E_{\bfk,\bfQ,\sigma}=E_{\bfk,\bfQ}-\sigma(h+\frac{\bfk\cdot\bfQ}{2m})$.

\section{Ginzburg-Landau functional}

Near the transition from a superfluid to normal state the order parameter is small.
We then  write the Ginzburg-Landau functional as $F = F_2 + F_4+...$, with
\beq
F_2 = \sum_{{\bf k},{\bf k}',n, m} \Delta_{\kk,\Q_m}(V^{-1})_{\kk, \kk'} \Delta^*_{\kk', \Q_m} - T\sum_{ \kk, n, m}  |\Delta_{\kk,\Q_m}|^2 G_+ (\kk+\frac{\Q_m}{2}, \om_n)
 G_- (-\kk +\frac{\Q_m}{2}, -\om_n);
 \eeq
 \beqa
 F_4 = \frac{T}{2}\sum_{  \kk, n, \{\Q_m\}}\Delta^*_{\kk,\Q_1}\Delta^*_{\kk,\Q_2}\Delta_{\kk,\Q_3}\Delta_{\kk,\Q_4}
  G_+ (\kk+\Q_1, \om_n)
 G_- (-\kk +\Q_3 -\Q_1, -\om_n) \nonumber \\
 \times G_+ (\kk+\Q_4, \om_n)
 G_- (-\kk, -\om_n) \delta (\Q_1 +\Q_2 - \Q_3 -\Q_4),
 \eeqa
where
\beq
G_{\pm} (\kk, \om_n) = \frac{1}{i\om_n - \xi_{\kk} \mp h},
\eeq
$\om_n = \pi T(2n + 1)$, $ \xi_{\kk} = \frac{k^2}{2m} - \mu$,
$\mu_{\uparrow, \downarrow}= \mu \pm h$, $\mu = \frac{k_F^2}{2m}$, and  $V^{-1}$ is the inverse matrix of $V_{{\bf k},{\bf k}'}$.
The  term $F_4$  is necessary to find the FFLO configuration corresponding to the energy minimum, whereas $F_2$
determines the tricritical point.

After the summation  over the  Matsubara frequencies we have
\beq
 F_2 = \sum_{{\bf k},{\bf k}', m} \Delta_{\kk,\Q_m}(V^{-1})_{\kk, \kk'} \Delta^*_{\kk', \Q_m} - \sum_{{\bf k}, m} \frac{|\Delta_{\kk,\Q_m}|^2}{2(\xi_{\kk+\Q_m/2}
 +\xi_{\kk-\Q_m/2})}\left(\tanh\frac{\xi_{\kk+\Q_m/2}+h}{2T}+ \tanh\frac{\xi_{\kk-\Q_m/2}-h}{2T}\right).
\label{f2}
 \eeq
The order parameter can be expanded in series over orbital angular momenta as
$\Delta_{\kk,\Q_m} = \sum_l \Delta_l (k, \Q_m) \exp(i \phi_{\kk} l)$,
and, similarly, $V_{\kk,\kk'} = \sum_l {V_l}_{k, k'} \exp[i (\phi_{\kk} -\phi_{\kk'})l]$, where $\phi_{\bf k}$ is the angle of the vector ${\bf k}$ with respect to the quantization axis.
In order to simplify the analysis of the functional we use an approximate solution for the order parameter, which is valid with a high accuracy.
For the $s$-wave pairing we have  $\Delta_{\kk, \Q_m} =  (V_{0,k,k_F}/V_{0,k_F,k_F}) \Delta_{\Q_m} \equiv \kappa_{k} \Delta_{\Q_m} $, where $\Delta_{\Q_m} = \Delta_0(k_F, \Q_m)$.
 Then we rewrite Eq.(\ref{f2}) as
 \beq
 F_2 = \sum_{m}\frac{|\Delta_{\Q_m}|^2}{V_{\mbox{eff}}} - \sum_{{\bf k}, m} \frac{\kappa_k^2 |\Delta_{\Q_m}|^2}{2(\xi_{\kk+\Q_m/2}
 +\xi_{\kk-\Q_m/2})}\left(\tanh\frac{\xi_{\kk+\Q_m/2}+h}{2T}+ \tanh\frac{\xi_{\kk-\Q_m/2}-h}{2T}\right) \equiv
\sum_m  \Omega_2(h, {\Q_m}, T)\Delta_{\Q_m}^2,
\label{f22}
 \eeq
 where  the effective interaction $V_{\mbox{eff}}$ is given by
 \beq
 \frac{1}{V_{\mbox{eff}}}= \sum_{{\bf k},{\bf k}'} \kappa_{\kk} (V^{-1})_{\kk, \kk'} \kappa_{\kk'}.
 \eeq
For the quantity $F_4$ we have:
 \beq
 F_4 = \frac{1}{2} \sum_{i,j}[ (2 - \delta_{\Q_i,\Q_j})| \Delta_{\Q_i}|^2 |\Delta_{\Q_j}|^2J(\phi_{\Q_i,\Q_j})
 +(1 -\delta_{\Q_i,\Q_j} -\delta_{\Q_i,-\Q_j}) \Delta_{\Q_i}\Delta_{-\Q_i}\Delta^*_{\Q_i}\Delta^*_{-\Q_i} \tilde{J}(\phi_{\Q_i,\Q_j})],
 \label{ff4}
 \eeq
 with
 \beq
 \rho(k_F)J(\phi_{\Q_i,\Q_j})= T\sum_{n, \kk}  G_- (\omega_n,
-\kk -\Q_i ) G_+(\omega_n,-\kk - \Q_j) G^2_+ (\omega_n,\kk ),
\eeq
\beq
 \tilde{J}(\phi_{\Q_i,\Q_j}) = T\sum_{n, \kk} G_-(\omega_n,-\kk) G_- (\omega_n,
-\kk -\Q_i -\Q_j) G_+(\omega_n,\kk + \Q_i) G_+ (\omega_n,\kk +\Q_j),
\eeq
where $\rho(k_F)$ is the density of states on the Fermi surface.
The sums  in  Eqs. (26) and (27) converge near the  Fermi surface and, therefore, we
substituted $\Delta_{\kk, \Q_m} = \Delta_{\Q_m}$ in Eq.(\ref{ff4}).

The line of the transition from the superfluid/FFLO to normal state is determined by the equation
\beq
\Omega_2(h_c, Q, T)= 0.    \label{critline}
\eeq
All $\Q_m$ have the same modulus $Q$ and, therefore, the quantity $\Omega_2$ is a function of $Q$.
For $h_c=0$ and $Q=0$ equation (\ref{critline}) gives the critical temperature $T_{c0}$ of the transition from the uniform superfluid to to normal state at zero field.
Putting $T=0$ in Eq.(\ref{critline}) we get the critical field $h_{c2}$ of the zero temperature transition from the FFLO to normal state.

In order to find the tricritical point  $T^*$ we expand the free energy at small $Q$:
\beq
\Omega_2 = A(T,  h) + B(T, h)Q^2 + C(T, h)Q^4.
\eeq
At $T=T^*$ the tricritical the coefficient $B$  changes a sign:
$$B(T^*, h^*) = \partial \Omega_2(h_c, Q, T_c)/\partial Q^2= 0;\,\,\,\,\,C > 0,B>0\,\, {\rm for}\,\, T> T^*.$$

\section{$(S + P)$-wave paired state}

The interlayer dipole-dipole interaction is long-range and for any orbital angular momentum
$l$  the leading part of the scattering (interaction) amplitude can be calculated in the Born approximation
(see Ref.~\cite{PS} and references therein). We have
\[
V_l(k, k') = \int d \phi_{\kk'} e^{i(\phi_{\kk'} - \phi_{\kk})} V_{\kk, \kk'},
\]
where the Fourier transform of the interlayer dipole-dipole potential is given by Eq.~(2)
of the main text. It is important that $V_l(k, k')$ is momentum dependent and partial amplitudes
for $|l|>0$ can be comparable with the $s$-wave amplitude $V_0$. The quantity $\rho (k) V_l (k, k)\lambda/r_*$ with $r_*=md^2$ being the dipole-dipole distance,
is a universal function of $k \lambda$. In Fig.1 we show the $k$-dependence  of this function  for
$l=0$ and $|l|=1$. Note that at larger $|l|$ it is much smaller  for realistic values of  $k \lambda$ ($k \lambda \lesssim 2$).
\begin{figure}[htb]
\includegraphics[width=8cm]{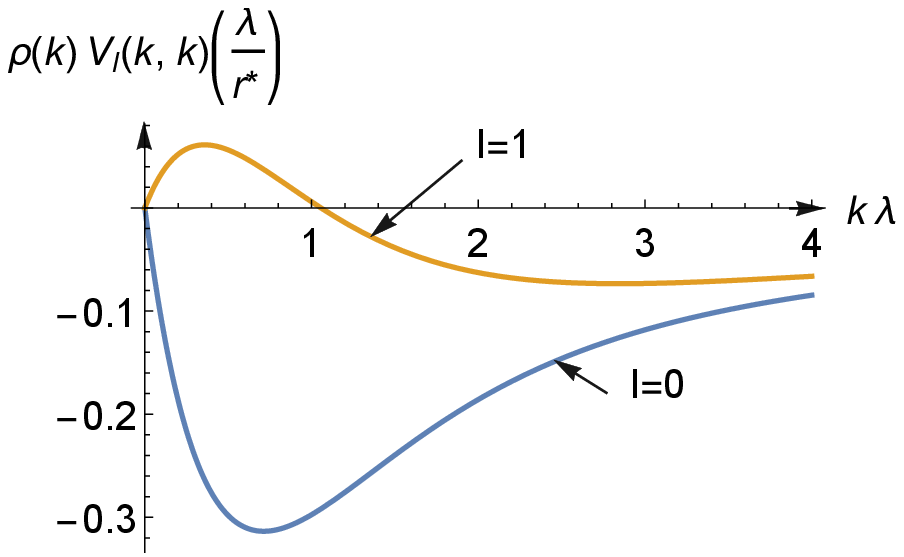}
\caption{The dependence of $\rho (k) V_l (k, k)\lambda/r_*$  on $k\lambda$ for $l=0$ and $l=1$.
}
\label{v01}
\end{figure}
For $k_F\lambda>1$ we have attraction in both $s$-wave and $p$-wave channels, i.e. $V_0(k_F,k_F)\equiv -V_s<0$ and $V_1(k_F,k_F)\equiv -V_p<0$. Therefore,  at a finite magnetic field $h$, which provides mixing between dipole-induced pairing with different orbital angular momenta, the order parameter can be a superposition of the $s$-wave and $p$-wave contributions:
\beq
\Delta_{\kk, \Q_m} = \Delta_s({\bf k},{\bf Q}_m) + \Delta_p({\bf k},{\bf Q}_m) \exp i \phi_{\kk}, \label{Deltasp}
\eeq
where $\Delta_s\equiv\Delta_{l=0}$ and $\Delta_p\equiv\Delta_{l=1}$.
In principle, one can think of  an  admixture  of the $p$-wave contribution to the order parameter also at
 $k_F \lambda <1$.  However, our numerics shows that this contribution is negligible.

In order to simplify the analysis we omit the k-dependencies of $\Delta_s$ and $\Delta_p$ and take their values  on the Fermi surface in Eq.~(22). Then,
instead of  Eq. (23) we have:
\begin{equation}     \label{F2Qm}
F_2=\sum_mF_2({\bf Q}_m),
\end{equation}
and
\beq
F_2({\bf Q}_m) = \frac{|\Delta_s({\bf Q}_m)|^2}{V_s} + \frac{|\Delta_p({\bf Q}_m)|^2}{V_p}  - T\sum_{\kk,n}  |\Delta_{\kk,\Q_m}|^2 G_+ (\kk+\frac{\Q_m}{2}, \om_n)
 G_- (-\kk +\frac{\Q_m}{2}, -\om_n).   \label{F2QmQ}
 \eeq
After the integration over $\kk$ we obtain
\beqa
\frac{F_2({\bf Q}_m)}{\rho(k_F)} =  \frac{|\Delta_s|^2}{\rho(k_F)V_s} + \frac{|\Delta_p|^2}{\rho(k_F) V_p} + 2\pi T {\rm Im}\sum_{n=0}^{\infty}
\frac{|\Delta_s|^2 + |\Delta_p|^2}{\sqrt{(i\bar{\om}_n)^2 - (h \bar{Q})^2}} \\ \nonumber
+2\pi T {\rm Im}\sum_{n=0}^{\infty} (\Delta_s \Delta_p^* + h.c.)\left[ \frac{1}{h \bar{Q}} -\frac{i\bar{\om}_n}{h \bar{Q}
\sqrt{(i\bar{\om}_n)^2 - (h \bar{Q})^2}}\right],
\eeqa
where $\bar{\om}_n = \om_n - i h$, and  $\bar Q = v_F Q/2h$ with $v_F$ being the  Fermi velocity.
We then transform  this equation to
\beqa
\frac{F_2({\bf Q}_m)}{\rho(k_F)} &=& \left[ \frac{|\Delta_s|^2}{\rho(k_F) V_s} -2\pi T {\rm Re}\sum_{n=0}^{\infty}  \frac{|\Delta_s|^2}{\bar{\om}_n}\right]
+ \left[\frac{|\Delta_p|^2}{\rho(k_F) V_p} -2\pi T {\rm Re}\sum_{n=0}^{\infty}  \frac{|\Delta_p|^2}{\bar{\om}_n}\right]  \nonumber \\
&-& \frac{1}{2}  {\rm Re} \int_{-\infty + i0}^{+\infty + i0} d\om \, \tanh\frac{\om}{2T}  (|\Delta_s|^2 + |\Delta_p|^2) \left[ \frac{1}{\sqrt{({\om}+ h)^2 - (h \bar{Q})^2}} -\frac{1}{{\om}+ h} \right]  \nonumber \\
&-&\frac{1}{2} {\rm Re}  \int_{-\infty + i0}^{+\infty + i0} d\om \, \tanh\frac{\om}{2T}      (\Delta_s \Delta_p^* + h.c.)\left[ \frac{1}{h \bar{Q}} -\frac{{\om} + h}{h \bar{Q}
\sqrt{({\om} + h)^2 - (h \bar{Q})^2}}\right].
\eeqa
\subsection{Zero temperature}
In the limit $T \to 0$  the Ginzburg-Landau functional can be calculated exactly:
\begin{eqnarray}
&\frac{F_2({\bf Q}_m)}{\rho(k_F)} = \left[|\Delta_s|^2 \ln \frac{2 h}{\Delta_0}\right] + \left[|\Delta_p|^2 \left( \frac{1}{\rho(k_F) V_p} - \frac{1}{\rho(k_F) V_s} +
\ln\frac{2 h}{\Delta_0} \right)\right] \nonumber \\
&+   (|\Delta_s|^2 + |\Delta_p|^2)\, {\rm Re}\left[ \ln (1 + \sqrt{1-\bar{Q}^2}) - \ln 2\right]
+ (\Delta_s \Delta_p^* + \Delta^*_s \Delta_p )\, {\rm Re} \left[ \frac{1 - \sqrt{1-\bar{Q}^2}}{\bar{Q}} \right], \label{F20}
\end{eqnarray}
where $\Delta_0$ is the gap  at $Q=0, T=0, h = 0$. The minimum of $F_2$ is achieved at  $\bar{Q} = 1$. The variation over $\Delta_s$, $\Delta_p$ gives
\begin{eqnarray}
&\Delta_s\ln \frac{h}{\Delta_0} - \Delta_p = 0,   \label{Dsp1} \\
&\left( \frac{1}{\rho(k_F) V_p} -\frac{1}{\rho(k_F) V_s} + \ln \frac{h}{\Delta_0} \right) \Delta_p - \Delta_s = 0.  \label{Dsp2}
\end{eqnarray}
The solution of these equations reads:
\begin{equation}     \label{Dspfin}
\Delta_p=\eta \Delta_s;\,\,\,\,\,\, h_c=\Delta_0  e^{\gamma},
\end{equation}
where
\begin{equation}     \label{gamma}
\eta=\frac{1}{2}\sqrt{\left( \frac{1}{\rho(k_F) V_p} -\frac{1}{\rho(k_F) V_s} \right)^2 +4\, \,} -\frac{1}{2}  \left( \frac{1}{\rho(k_F) V_p} -\frac{1}{\rho(k_F) V_s} \right).
\end{equation}

In the limit $ \rho(k_F) V_p \ll \rho(k_F) V_s \ll 1$  we have:
\beq
\eta=\rho(k_F)V_p\ll 1;\,\,\,\,\,\,h_c\simeq\Delta_0;\,\,\,\,\,\Delta_p\ll \Delta_s. \label{hcD0}
\eeq
Thus, in this limit, like for $k_F\lambda<1$ where the interaction in the $p$-wave channel is repulsive and we have only the $s$-wave pairing, or like in the case of contact $s$-wave attraction, the critical field $h_{c_1}$
is equal to $\Delta_0$. However, for $k_F\lambda$ significantly exceeding unity the ratio $h_{c_1}/\Delta_0$ noticeably increases. Already for $k_F\lambda\simeq 2.2$ and $r_*$ approaching $\lambda$ we have $h_{c_1}/\Delta_0\simeq 1.12$.

Note that due to the long-range character of the dipole-dipole interaction the $k$-dependence of $\Delta_s$ and $\Delta_p$ can play a role (see below). In particular, it slightly reduces the ratio $h_c/\Delta_0$. Moreover, for $k_F\lambda<1$ where we have only the $s$-wave pairing, the numerical solution of the Gor'kov equations (see below) shows that the critical field becomes smaller than $\Delta_0$  (see Fig.~2).
\begin{figure}[htb]
\includegraphics[width=8cm]{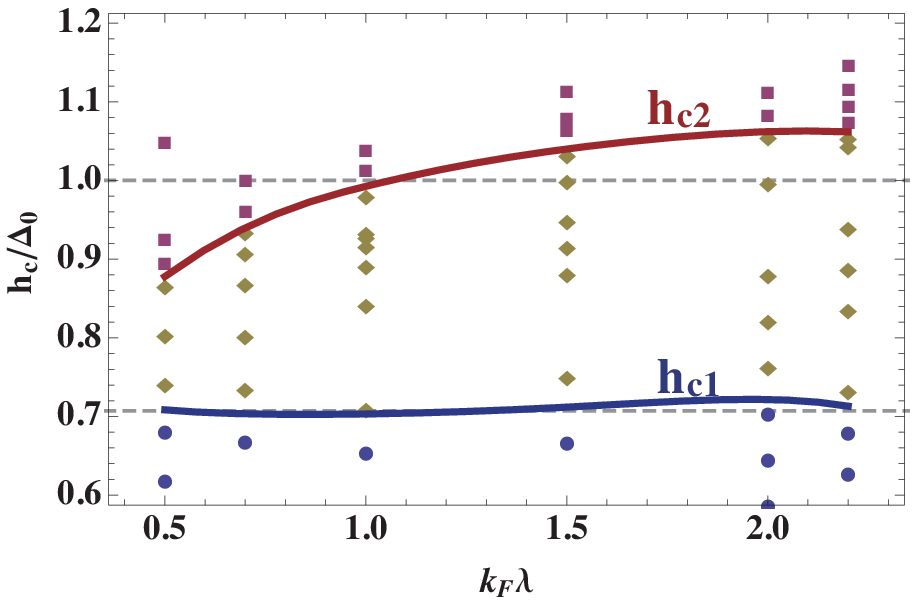}
\caption{Critical fields $h_{c1}$ and $h_{c2}$ at $T=0$ in units of $\Delta_0$ versus the parameter $k_F \lambda$ for $k_Fr_*=0.7$. The dark blue curve is $h_{c1}$, and the dark red curve $h_{c2}$. The dashed lines show $h_{c1}$ and $h_{c2}$ for the case of contact interaction. The blue, light brown and red dots are the points where our numerical solution of the self-consistent Gor'kov equations (11)-(15) gives uniform superfluid, FFLO, and normal phase,  respectively.
}
\label{0Tphase}
\end{figure}

\subsection{The vicinity of  the tricritical point.}
In the vicinity of the tricritical point $T^*, h^*$ the momenta $Q_m$ are small. It is then %convinient
convenient to first write Eq.(\ref{F2QmQ})
without an explicit integration over the angle and perform the integration after expanding $F_2$ in powers of $Q_m$. We thus have:
\beqa
\frac{F_2}{\rho(k_F)} = - |\Delta_s|^2  \ln \frac{T_{0,s}}{T} + |\Delta_s|^2  \int \frac{d\phi}{2\pi}
 {\rm Re} \left[\Psi \left( \frac{1}{2} + i\frac{2 h + v_F Q  \cos \phi }{4\pi T}\right) -\Psi \left( \frac{1}{2}\right)\right] \nonumber \\
 - |\Delta_p|^2  \ln \frac{T_{0,p}}{T} + |\Delta_p|^2  \int \frac{d\phi}{2\pi} {\rm Re}
  \left[\Psi\left(\frac{1}{2} + i\frac{2 h + v_F Q  \cos \phi }{4\pi T}\right) -\Psi\left(\frac{1}{2}\right)\right] \nonumber \\
 +(\Delta_s^* \Delta_p + \Delta_s \Delta_p^*)  \int \frac{ \cos \phi \, d\phi}{2\pi} {\rm Re}
  \left[\Psi\left(\frac{1}{2} + i\frac{2 h + v_F Q  \cos \phi }{4\pi T}\right) \right], \label{F2Q}
 \eeqa
 %or
 %\beqa
 %\frac{F_2}{N_0} =  (|\Delta_s|^2 + |\Delta_p|^2 )\left[ - \ln \frac{T_{0,s}}{T} +   \int \frac{d\phi}{2\pi}
 %{\rm Re} \left[\Psi \left( \frac{1}{2} + i\frac{2 h + v_F q  \cos \phi }{4\pi T}\right) -\Psi \left( \frac{1}{2}\right)\right]\right] \nonumber \\
% + |\Delta_p|^2  \ln \frac{T_{0,s}}{T_{0,p}} +
 %+(\Delta_s^* \Delta_p + \Delta_s \Delta_p^*)  \int \frac{ \cos \phi \, d\phi}{2\pi} {\rm Re}
 % \left[\Psi\left(\frac{1}{2} + i\frac{2 h + v_F q  \cos \phi }{4\pi T}\right) \right],
 %\eeqa
  where $Q\equiv Q_m$, the quantities $T_{0,s}$ and $T_{0,p}$ are transition temperatures at $h=0$ for the purely $s$-wave and purely $p$-wave superfluidity, and
  $\Psi (x)= \Gamma'(x)/\Gamma (x) $ is the digamma  function. The limiting case $T\rightarrow 0$ described by equation (\ref{F20}) follows from Eq.(\ref{F2Q}) by using the asymptotic relation
$\Psi (1/2 + i x)= i \pi/2 + \ln x$ for $x\rightarrow +\infty$.

At the tricritical point one has $F_2=0$, which means that first of all the zero order term of the expansion of $F_2$ (\ref{F2Q}) in powers of $Q$ is equal to zero. In the case of $\Delta_p=0, T_{0,p}=0$ the third line of (\ref{F2Q}) is zero. The expansion of the first two lines gives the zero and second order terms. Therefore, the latter should be also equal to zero. This leads to the known result for $T^*, h^*$ \cite{B,S}:
\beq
-  \ln \frac{T_{0,s}}{T^*} +
 {\rm Re} \left[\Psi \left( \frac{1}{2} + i\frac{2 h^*  }{4\pi T^*}\right) -\Psi \left( \frac{1}{2}\right)\right] =0; \label{tricrs1}
\eeq

\beq
 {\rm Re} \left[\Psi^{''} \left( \frac{1}{2} + i\frac{2 h^*  }{4\pi T^*}\right)\right] =0; \label{tricrs2}
 \eeq

\beq
 \frac{h^*}{2 \pi T^*} \simeq 0.3, \quad T^* = 0.56 T_{0,s}. \label{tricrs}
\eeq

For a finite $\Delta_p$ we expand $F_2$ (\ref{F2Q}) in powers of $Q$ up to the fourth order terms. After performing the integration over the angle $\phi$ we then obtain:
\beq
\frac{F_2}{\rho(k_F)} = A (|\Delta_s|^2 + |\Delta_p|^2)  + |\Delta_p|^2  \ln \frac{T_{0,s}}{T_{0,p}} + ( \alpha Q^2 + \gamma Q^4)  ( |\Delta_p|^2 + |\Delta_s|^2) +
(\Delta_s^* \Delta_p + \Delta_s \Delta_p^*) \beta Q,      \label{F2Qexp1}
\eeq
where $A$ is given by equation (\ref{tricrs1}) with $T^*$ replaced with $T$. Other coefficients in Eq.(\ref{F2Qexp1}) are given by
\beq      \label{coeff}
\quad \alpha  =  -\frac{P\xi^2}{16}{\rm Re}\Psi''\left(\frac{1}{2}+\frac{ih}{2\pi T}\right), \quad
 \gamma =\frac{\xi^4}{128}{\rm Re} \Psi^{IV}\left(\frac{1}{2}+\frac{ih}{2\pi T}\right), \quad  \beta =   -\frac{\xi}{4}{\rm Im} \Psi'\left(\frac{1}{2}+\frac{ih}{2\pi T}\right), \quad   \xi = \frac{v_F}{2 \pi T} .
\eeq

Minimizing $F_2$ (\ref{F2Qexp1}) with respect to $\Delta_{\p}$ we find:
\beq    \label{Deltap}
\Delta_p  =- \frac{\Delta_s \beta Q}{\ln \frac{T_{0,s}}{T_{0,p}}+ A + \alpha Q^2+ \gamma Q^4}.
\eeq
The functional $F_2$ then acquires the form
\beq     \label{F2QQ}
F_2=|\Delta_s|^2 \left[ A +(\alpha Q^2 + \gamma Q^4) -\frac{\beta^2 Q^2}{\ln \frac{T_{0,s}}{T_{0,p}}+A }+\frac{\alpha\beta^2 Q^4}{\left(\ln \frac{T_{0,s}}{T_{0,p}}+A\right)^2 } \right],
\eeq
where we omitted terms containing the power of $Q$ that is higher than $4$. At $T=T^*$ the term independent of $Q$ in Eq.(\ref{F2QQ}) should be equal to zero, i.e. $A=0$ and the first equation for finding the point
$T^*, h^*$ is Eq.(\ref{tricrs1}). The second equation is obtained from the condition that the coefficient in front of $Q^2$ changes sign at this point (negative for $T<T^*$ and positive for $T>T^*$), so that it is zero at $T=T^*$. Since $A=0$, we have
$\alpha=\beta^2/\ln{(T_{0,s}/T_{0,p})}$ and using relations (\ref{coeff}) we find:
 \beq        \label{tricrsp}
 -{\rm Re}\left[\Psi''  \left( \frac{1}{2} + i\frac{h^*  }{2\pi T^*}\right)\right] = \frac{\left({\rm Im} \Psi' \left( \frac{1}{2} + i\frac{h^*  }{2\pi T^*}\right)   \right)^2}{\ln{(T_{0,s}}/T_{0,p})},
 \eeq
which gives Eq.({\ref{tricrs2}) in the limit $T_{0,p}  \to 0$. The quantities $\ln(T_{0,s}/T_{0,p})$, $h^*/2\pi T^*$, and $T^*/T_{0,s}$, are universal functions of $k_F\lambda$ and $k_Fr_*$. Their dependence on $k_F\lambda$ at $r_*/\lambda=1$ is shown in Fig.3, Fig4, and Fig5.
\begin{figure}[htb]
\includegraphics[width=8cm]{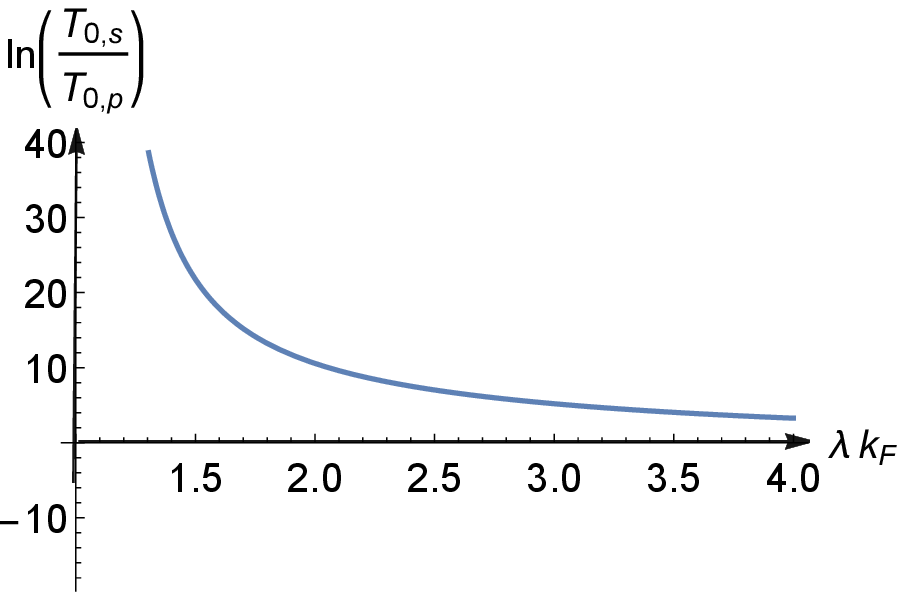}
\caption{The quantity $\ln(T_{0,s}/T_{0,p})$  as a function of $k_F \lambda$ for $r_*=\lambda$.
}
\label{}
\end{figure}
\begin{figure}[htb]
\includegraphics[width=8cm]{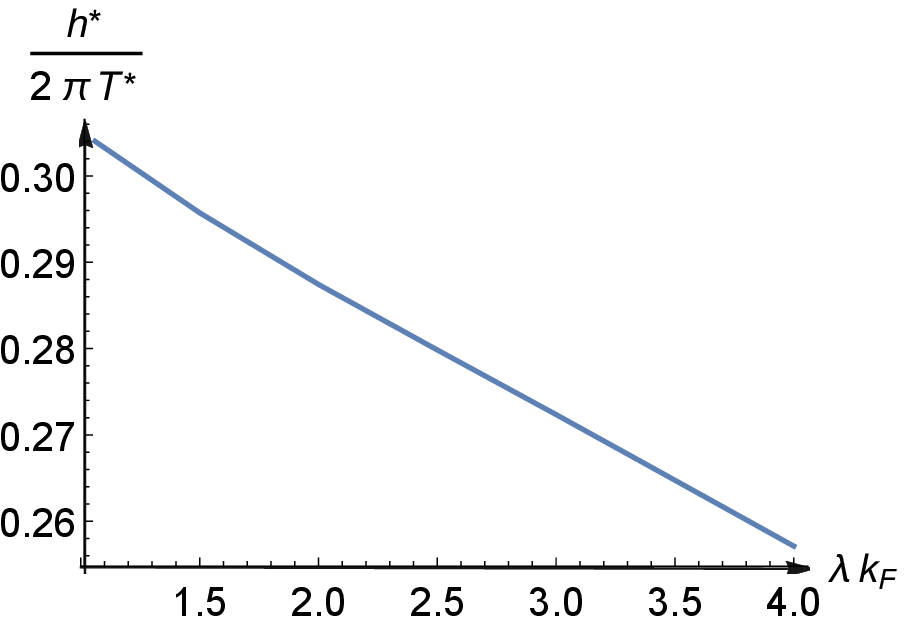}
\caption{The ratio $h^*/2\pi T^*$  as a function of $k_F \lambda$ for $r_*=\lambda$.
}
\label{ff2}
\end{figure}
\begin{figure}[htb]
\includegraphics[width=8cm]{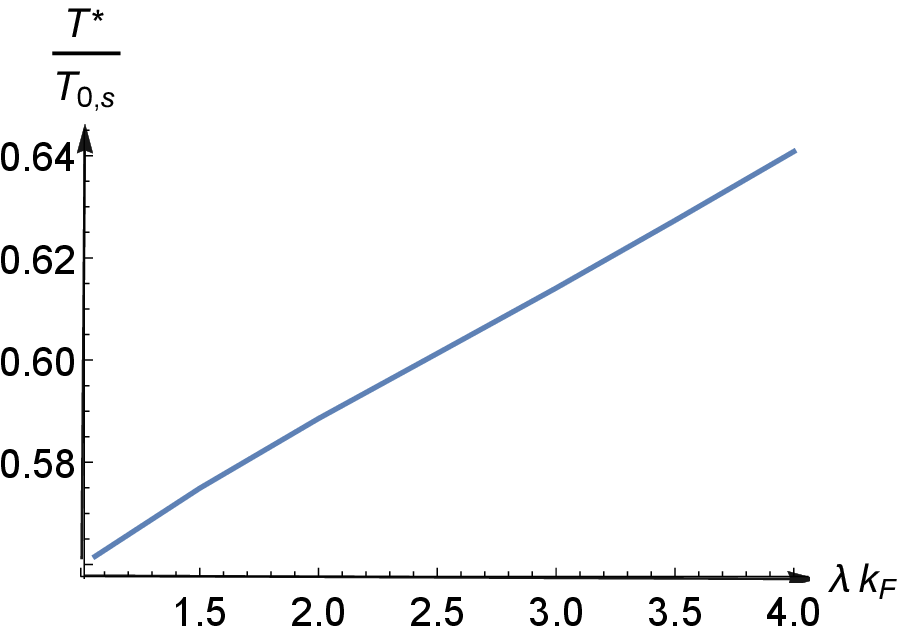}
\caption{The  dependence of the dimensionless tricritical  temperature  $T^*/T_{0,s}$ on  $k_F \lambda$ for $r_*=\lambda$.
}
\label{}
\end{figure}

Since the gap equation  is linear, any linear combination
of plane waves (any number of $\Q_m$)   is
allowed. In order to find the most energetically
favorable combination, one has to calculate the contribution
$F_4$ to the free energy and put  $T= T^*$ in this contribution.

For finding the dependence $Q(T)$ near the tricritical point we minimize $F_2$ with respect to $Q^2$ (the contribution proportional to $Q^4$ is also contained in $F_4$, but it is proportional to $|\Delta_s|^4$ and can be omitted).
The coefficient in front of $Q^2$ is proportional to $(T^*-T)$ and we have:
\begin{equation}
\frac{dF_2}{dQ^2}=-(T^*-T)B_1+B_2Q^2=0,
\end{equation}
where
\beq
B_1=\frac{d\alpha}{dT}-\frac{1}{\ln(T_{0,s}/T_{0,p})}\frac{d\beta^2}{dT}+\frac{\beta^2}{\ln^2(T_{0,s}/T_{0,p})}\frac{dA}{dT}.
\eeq
\beq
B_2=2\left(\gamma+\frac{\alpha\beta^2}{\ln^2(T_{0,s}/T_{0,p})}\right).
\eeq
This yields:
\beq
Q=\sqrt{(T^*-T)B_1/B_2}.  \label{QB}
\eeq
The result of Eq.(\ref{QB}) is very close to that obtained by numerical solution of the Gor'kov equations and displayed in Fig.8 (see below).

%%%%%%%%%%%%%%%%%%%%%%%%%%%%
\section{Description of  the numerical procedure}

In this section, we first describe the procedure of solving the self-consistent Gor'kov equations (13)-(15) together with self-consistent gap equations (11), (12) and minimizing the free energy for a giving set of parameters. We then provide some mode details about the numerical results for the phase diagram.

\subsection{Solving the self-consistent Gor'kov equations and minimizing the free energy}
%at a specific set of the parameters, $(r_*,\lambda,h,T,Q,N_Q)$.}

Due to the 2D rotational symmetry, we can first decompose the self-consistent gap equations (11) and (12)  into orbital angular momentum channels:
\begin{eqnarray}
&\Delta_{{\bf k},{\bf Q}_m}=\sum_l\Delta_l(k,{\bf Q}_m)\exp(i\phi_kl);      \label{Deltal} \\
&\Delta_l(k,\Q_m)=-\int^{\infty}_0 \frac{k'dk'}{2\pi}V_{l,k,k'}\sum_{l'}\Delta_{l'}(k',\Q_m)P_{l-l'}(k',\Q_m,h,T), 
\label{gappartial}
\end{eqnarray}
and similar equations hold for $\Delta^{\dagger}_{{\bf k},{\bf Q}_m}$.
Here $P_{l}(k,\Q_m,h,T)\equiv\int^{2\pi}_0 P(\kk,\Q_m,h,T)e^{-il\phi}\frac{d\phi}{2\pi}$, and $P(\kk,\Q_m,h,T)\equiv T\sum_n F_{\uparrow,\downarrow}(\kk+\Q_m/2,\kk-\Q_m/2;i\omega_n)/\Delta_{\kk,\Q_m}$, with 
the anomalous Green function $F_{\uparrow,\downarrow}(\kk,\kk';i\omega_n)$ from Eq.(\ref{Fd}), and 
$V_{l,k,k'}\equiv\int_0^{2\pi}\frac{d\phi_\kk}{2\pi}V_{\kk,\kk'}\,e^{-i(\phi_\kk-\phi_{\kk'})l}$ is the $l$-th angular momentum component of the interlayer dipolar interaction. Note that we have $\Q_1=0$ for the uniform superfluid, 
$\Q_1=Q\hat{x}$ for the plane wave FFLO, $\Q_{1,2}=\pm Q\hat{x}$ for the stripe FFLO, 
and $\Q_n=Q(\cos(n-1)\frac{2\pi}{3}\hat{x}+\sin(n-1)\frac{2\pi}{3}\hat{y})$ for the triangular FFLO.

The gap equation (\ref{gappartial}) is solved numerically by a standard iteration method for a given set of $\Q_m$ at given temperature $T$ and population imbalance $h$. The functions $F$ and $F^{\dagger}$ are taken from   equations (13)-(15) in which $\Delta_{{\bf k},{\bf Q}_m}$ is described by equations (\ref{Deltal}) and (\ref{gappartial}). Choosing the momentum cut-off much larger than the Fermi momentum (specifically $k_{cut}\geq 10k_F$ in all cases), the trial gap function $\Delta_l(k)$ converges such that the error, $\delta\Delta_l(k)/\Delta_l(k)$, is within $10^{-6}$ at every $k$-point. As an example, in Fig. \ref{gap function} we show the calculated gap functions for the plane wave FFLO at $k_F\lambda=0.5$ and $k_F\lambda=2.2$, respectively. 
We can see that the gap function has a significant $k$-dependence in both $s$-wave and $p$-wave channels, especially when $k_F\lambda$ increases. It is also seen that for fairly small $k_F\lambda$ the $p$-wave contribution is practically negligible.   

%--------------
\begin{figure}[htb]
\includegraphics[width=15cm]{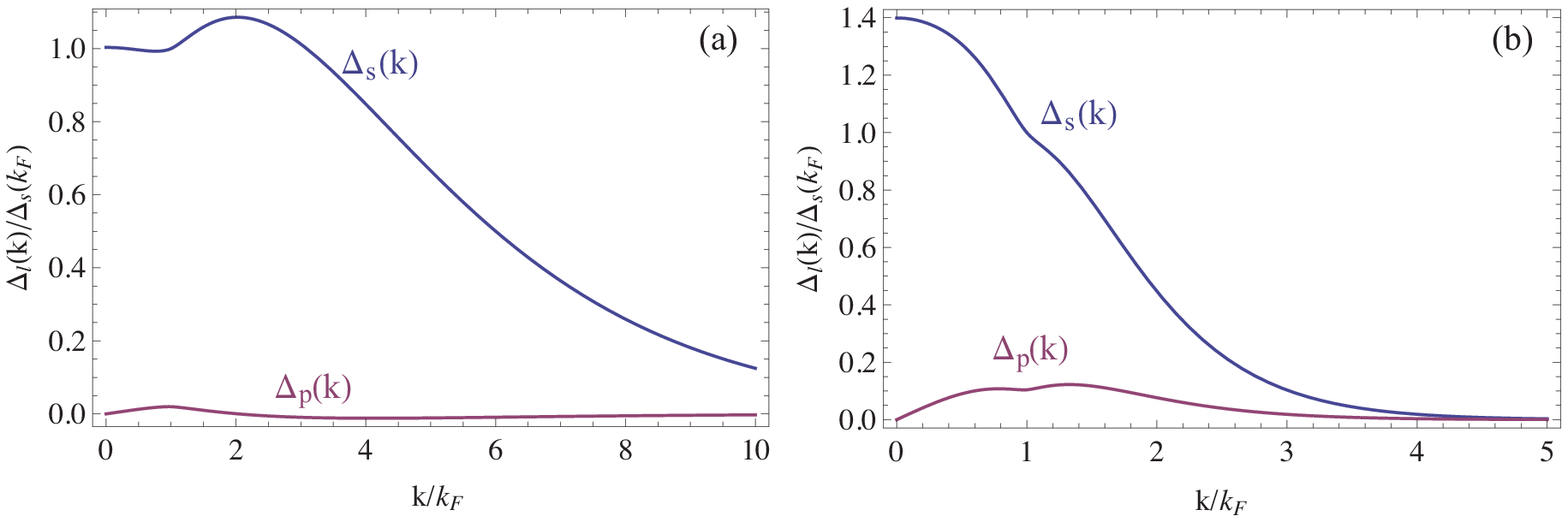}
\caption{Zero temperature gap function of a plane wave FFLO for both $s$-wave and $p$-wave components. In (a) $k_F\lambda=0.5$ and $h/\Delta_0=0.9$, and in (b) $k_F\lambda=2.2$ and $h/\Delta_0=1.03$. The results for the stripe and triangular phases are similar. 
}
\label{gap function}
\end{figure}
%---------------

In order to efficiently solve the gap equation at a finite imbalance ($h\neq 0$), 
the iteration starts with a trial gap function, $\Delta_l(k)\propto V_{l,k,k_F}$, where $l=0,1$ for the $s$-wave and $p$-wave components, respectively. For convenience we define $\Delta_0\equiv\Delta_s(k_F,Q=0)$ for $T=h=0$ as the modulus of the uniform BCS superfluid gap. Once the gap function is obtained, we substitute it into Eqs.~(\ref{exi})-(\ref{E}) and calculate the total free energy. The equilibrium phase is determined by minimizing the free energy with respect to different sets of $(Q,N_Q)$, where $N_Q=0,1,2,3$ for uniform superfluid, plane wave FFLO, stripe FFLO, and triangular FFLO, respectively. The obtained results are referred to as the phase for a given set of parameters, $(r_*,\lambda,h,T)$. 

\subsection{Numerical results in detail}

\subsubsection{FFLO at zero-temperature}

At zero temperature we obtain three phases as the population imbalance (magnetic field) increases. The ground state is a uniform superfluid for $h<h_{c1}$ and becomes the FFLO state for $h>h_{c1}$. The phase transition is known to be of the first order as the order parameters in both regimes are finite at the phase boundary. As the imbalance is further increased, the FFLO order parameter decreases to zero when $h>h_{c2}$, so that the ground state becomes normal through a second order phase transition. In this sense the results are similar to those obtained for contact interactions.  

In our case of dipolar interaction, we find that the energy of the stripe FFLO is always lower than that of the plane wave and triangular FFLO states. The difference in energies of the stripe phase and the two other candidates is of the order of $10^{-6}\epsilon_F$ or higher, which is by an order of magnitude larger than the numerical uncertainty of our calculation. The energy difference becomes comparable with the uncertainty of the calculation only in a tiny region where $|h-h_{c2}|/h_{c2}$ is smaller than a few percent. 

In the system with contact interactions at $T=0$, the phase boundaries are predicted to be universal: $h_{c1}/\Delta_0=1/\sqrt{2}$ and $h_{c2}/\Delta_0=1$ (see Ref.\cite{Bulaevskii}). In contrast, in our case of interlayer dipolar interaction they are non-universal and depend on the the parameter $k_Fr_*$ and especially on $k_F\lambda$. This is shown in Fig.\ref{0Tphase}, where one sees that for $k_F\lambda>1$ the field $h_{c2}$ is significantly larger than $\Delta_0$, so that the FFLO region becomes wider than in the case of contact interactions. This is because for such $k_F\lambda$ the $p$-wave interaction becomes also attractive and Cooper pairs are composed of both $s$-wave and $p$-wave contributions. This makes the modulus of the order parameter larger and requires a higher field $h_{c2}$ to destroy superfluidity. On the other hand, for $k_F\lambda<1$ the $p$-wave interaction is repulsive and the superfluid pairing is practically $s$-wave. As a result, the FFLO regime is even suppressed, i.e. $h_{c2}/\Delta_0<1$. The value of $h_{c1}/\Delta_0$ does not change much compared to the contact interaction case, since the associated  phase transition is of the first order. 
\subsubsection{FFLO at finite temperature}

At finite temperatures we numerically confirm that the critical temperature of the uniform superfluid is $T_{0s}\approx 0.57\Delta_0$ at $h=0$, like in the case of contact interactions. The finite temperature phases are determined by minimizing the total free energy as described above.
\vspace{2mm}
\begin{figure}[htb]
\includegraphics[width=17cm]{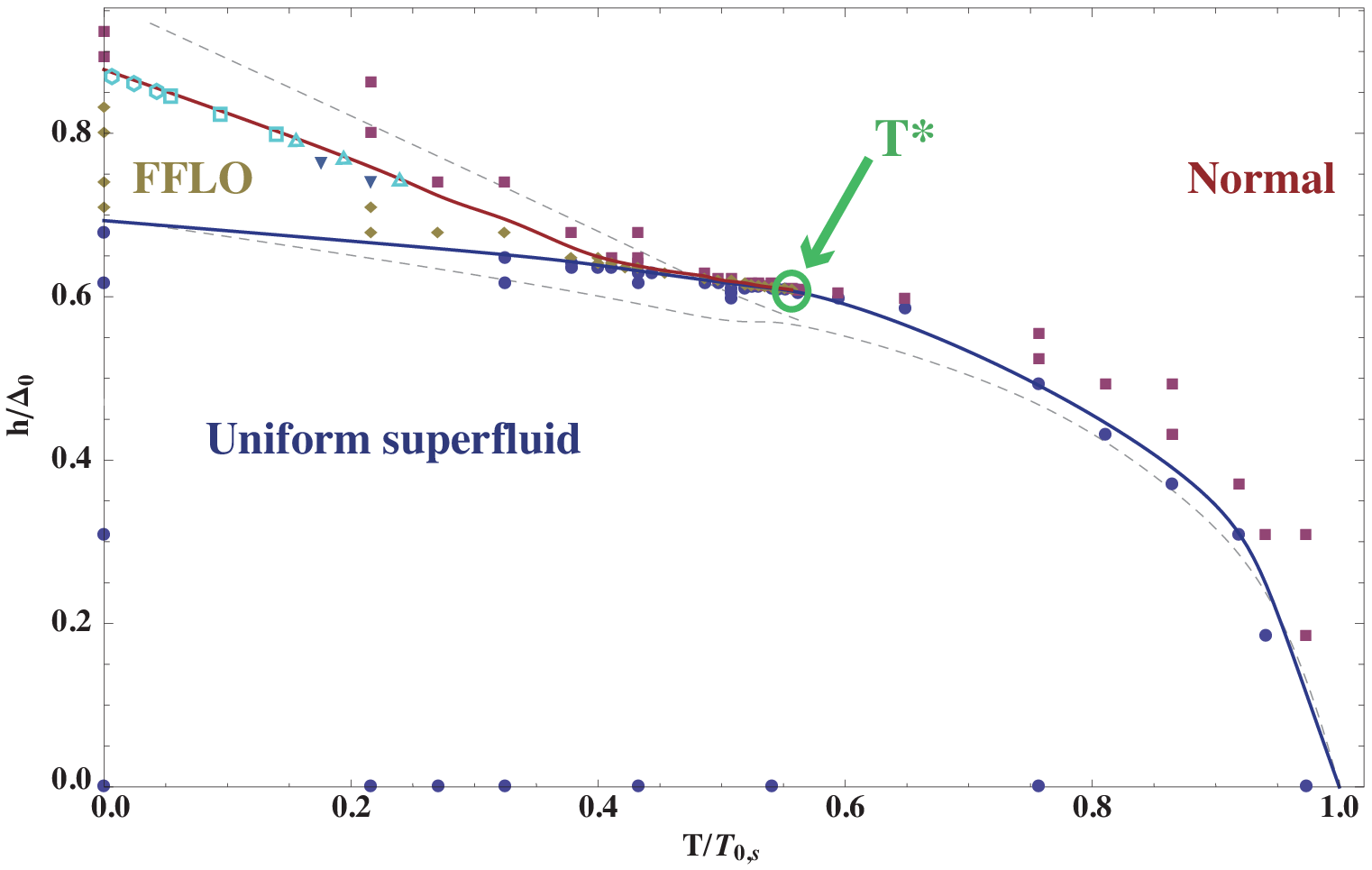}
\caption{Finite temperature phase diagram in terms of $T/T_{0s}$  and the imbalance
 $h/\Delta_0$  at
 $k_F\lambda =0.5$ and $r_*=0.5 \lambda$. %The blue, light brown, and red dots 
The blue circles, light brown diamonds, and red squares are the points where our numerical solution of the self-consistent Gor'kov equations (11)-(15) (or, very close to the transition line, the calculation from Ginzburg-Landau-based equations (28)-(32)) gives uniform superfluid, stripe FFLO, and normal phase,  respectively. The cyan triangles, squares, and hexagons indicate the points where the calculation from Ginzburg-Landau-based equations (28)-(32) gives triangular, square, and hexagonal FFLO states. The %filles
filled triangles show the points where the triangular FFLO is obtained from the solution of the self-consistent Gor'kov equations.
}
\label{fTphase05}
\end{figure}
%---------------
\begin{figure}[htb]
\includegraphics[width=17cm]{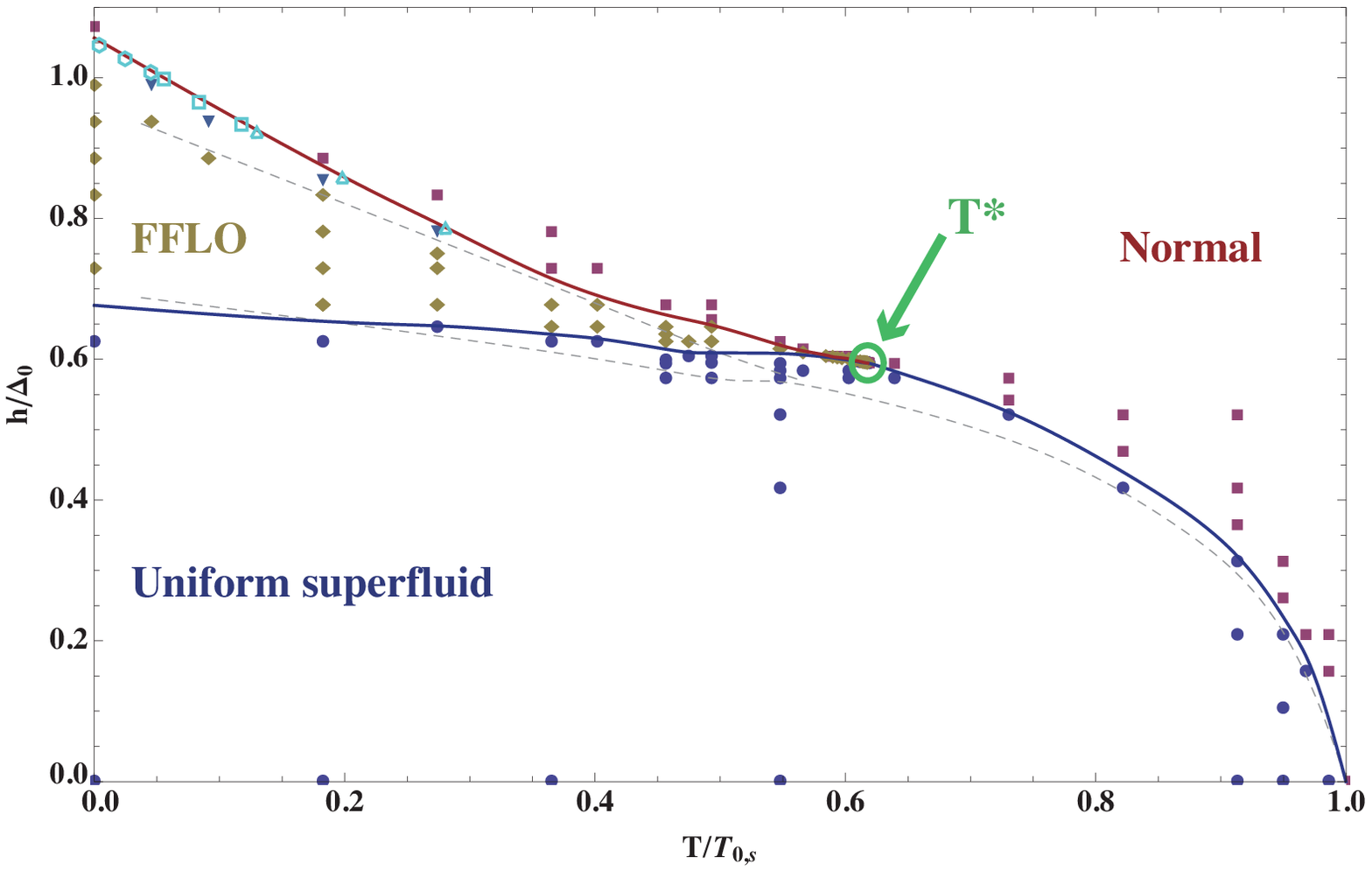}
\caption{Finite temperature phase diagram in terms of   $T/T_{0s}$  and the imbalance
 $h/\Delta_0$  at
 $k_F\lambda =2.2$ and $r_*=\lambda$. %The blue, light brown, and red dots 
The blue circles, light brown diamonds, and red squares are the points where our numerical solution of the self-consistent Gor'kov equations (11)-(15) (or, very close to the transition line, the calculation from Ginzburg-Landau-based equations (28)-(32)) gives uniform superfluid, stripe FFLO, and normal phase,  respectively. The cyan triangles, squares, and hexagons indicate the points where the calculation from Ginzburg-Landau-based equations (28)-(32) gives triangular, square, and hexagonal FFLO states. The %filles
filled triangles show the points where the triangular FFLO is obtained from the solution of the self-consistent Gor'kov equations.
}
\label{fTphase22}
\end{figure}
%---------------
In Fig. \ref{fTphase05} and Fig. \ref{fTphase22} we show the numerically obtained finite temperature phase diagrams at $k_F\lambda=0.5$ and $k_F\lambda=2.2$, respectively. As the temperature increases from $T=0$, the fields $h_{c1}$ (the blue curves) and $h_{c2}$ (the dark red curves) both decrease and merge with each other at the tricritical temperature $T^*$. For comparison, we also show the phase boundaries of the contact interaction case (gray curves).

As well as the zero-%temperarture
temperature critical field $h_{c2}/\Delta_0$ discussed above, the tricitical temperature $T^*/T_{0s}$ is enlarged as the p-wave interaction becomes attractive, in agreement with the analytical Ginzburg-Landau calculations presented above. 
Note that in Fig. \ref{fTphase05} ($k_F\lambda=0.5$) we have $T^*/T_{0s}\approx0.56$, which is the same as in the contact interaction case, while at $k_F\lambda=2.2$ the tricritical temperature is $T^*/T_{0s}\approx0.62$ (Fig. \ref{fTphase22}), showing a strong effect of the p-wave attraction. 

%------------------
\begin{figure}[htb]
\includegraphics[width=8cm]{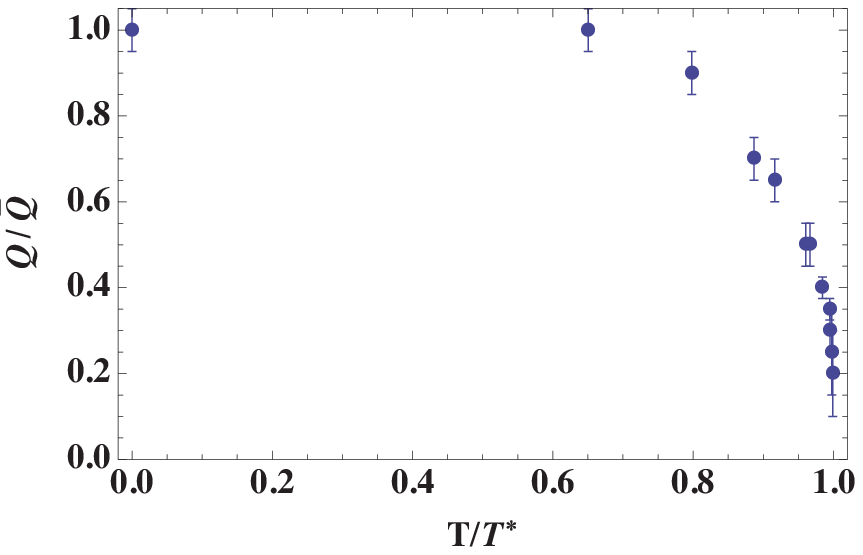}
\caption{The optimal $Q/\bar{Q}$ at $h$ close to $h_{c2}$ in terms of $T/T_*$ 
at $k_F\lambda =2.2$ and $r_*=\lambda$, where $\bar{Q}\equiv2mh/k_F$.
}
\label{fTQ}
\end{figure}

The equilibrium FFLO state emerges as the stripe phase, except for a narrow region of $h$ near $h_{c2}$, where the difference in energies of the considered FFLO structures is comparable with the uncertainty of our numerical calculations from the self-consistent Gor'kov equations. In the temperature interval from %$0.02T^*$ to $0.3T^*$
$0.02T_{0,s}$ to $0.3T_{0,s}$ this is the case for $|h-h_{c2}|/h_{c2}\lesssim 0.05$. For such $h$ our calculations based on the Ginzburg-Landau approach indicate that the equilibrium state is a triangular FFLO (it is also recovered from the self-consistent Gor'kov equations at $h\simeq 0.95h_{c2}$), which with decreasing temperature becomes a square and then hexagonal FFLO (see Fig.7 and Fig.8). This sequence of FFLO states is similar to that found in the case of contact interactions \cite{S,Combescot}.

Finally, in Fig. \ref{fTQ}, we show how the numerically obtained optimal $Q$ (for the state of the lowest free energy) decreases with increasing $T$ along the $h_{c2}$ transition line and eventually approaches zero as $T\rightarrow T^*$. The way it approaches zero is the same as the one predicted by the Ginzburg-Landau theory in Eq.(\ref{QB}), i.e. $Q\propto\sqrt{T^*-T}$.